\documentclass[11pt]{article}
\usepackage{amsfonts}
\usepackage{amssymb}          
\usepackage{a4}
\usepackage{graphics}

\newcommand{\bref}[1]{(\ref{#1})}

\newcommand{\bR}{\mathbb{R}}
\newcommand{\bC}{\mathbb{C}} 
\newcommand{\bT}{\mathbb{T}}
\newcommand{\bZ}{\mathbb{Z}}
\newcommand{\tfrac}[2]{{\textstyle{\frac{#1}{#2}}}}
\newcommand{\frt}{\tfrac{1}{\sqrt{2}}}
\newcommand{\half}{\tfrac{1}{2}}
\newcommand{\quarter}{\tfrac{1}{4}}
\newcommand{\imu}{{\rm i}\,}
\newcommand{\edth}{\eth}
\newcommand{\edthbar}{\bar{\eth}}
\newcommand{\mbar}{\bar{m}}
\newcommand{\cA}{\mathcal{A}}
\newcommand{\cD}{\mathcal{D}}
\newcommand{\cN}{\mathcal{N}}

\newcommand{\cL}{\mathcal{L}}
\newcommand{\cK}{\mathcal{K}}
\newcommand{\cB}{\mathcal{B}}
\newcommand{\scri}{\mathcal{I}}

\newcommand{\mschw}{\textsc{M}}
\newcommand{\scF}{\textsc{F}}

\newcommand{\Qp}{Q^{+}}
\newcommand{\Qm}{Q^{-}}
\newcommand{\Qpm}{Q^{\pm}}

\newcommand{\Qmbar}{\bar{Q}^{-}}
\renewcommand{\div}{{\mbox{div}}}
\newcommand{\curl}{{\mbox{curl}}}
\renewcommand{\Re}{{\mbox{Re}}}
\renewcommand{\Im}{{\mbox{Im}}}

\begin{document}

\begin{titlepage}

\hspace*{11cm} 
\parbox[t]{3cm}
{ \small gr-qc/9904045 \hfill \\ \small ISE RR 99/110 }

\vspace*{1cm}
\begin{center} \Large
   Einstein equations in the null quasi-spherical gauge III: \\
         numerical algorithms
\end{center}
\vspace*{1cm}
\begin{center} \large
   Robert Bartnik\footnote{\tt Bartnik@ise.canberra.edu.au} 
        and Andrew H.~Norton\footnote{\tt AndrewN@ise.canberra.edu.au}\\
School of Mathematics and Statistics\\
University of Canberra\\
ACT 2601, Australia\\
\end{center}
\vspace*{1cm}

\begin{abstract}
We describe numerical techniques used in our construction of a 4th 
order in time evolution for the full Einstein equations, and assess 
the accuracy of some representative solutions.  The scheme employs 
several novel geometric and numerical techniques, including a 
geometrically invariant coordinate gauge, which leads to a 
characteristic-transport formulation of the underlying hyperbolic 
system, combined with a ``method of lines'' evolution; a convolution 
spline for radial interpolation, regridding, differentiation and noise 
suppression; representations using spin-weighted spherical harmonics; 
and a spectral preconditioner for solving a class of 1st order 
elliptic systems on $S^{2}$.  Initial data for the evolution is 
unconstrained, subject only to a mild size condition.  For 
sample initial data of ``intermediate'' strength ($19\%$ of the
total mass in gravitational energy), the code is accurate to 1 part in 
$10^{5}$, until null time $z=55\mschw$ when the coordinate condition 
breaks down.
\end{abstract}

\vfill

This project has been supported by 
Australian Research Council grants A69330046 and A69802586.

\end{titlepage}

\section{Introduction}

The Einstein equations present difficulties of size and complexity 
somewhat greater than those normally encountered in scientific 
computation.  A numerical simulation must confront a range of 
theoretical and numerical challenges, starting with the theoretical 
problem of finding a well-posed and geometrically natural reduction of 
the full system of Einstein equations.  Numerical algorithms are then 
needed to control the various facets of the reduced system, 
efficiently, since very large data structures are inevitable when 
modelling fully 3+1-dimensional spacetimes.  Finally there are the 
twin theoretical and practical problems of understanding the nature of 
the solution represented by the data, and of certifying its 
reliability.

In \cite{Bartnik97a} we presented a new coordinate formulation for the 
vacuum Einstein equations, based on a characteristic (null) coordinate 
and a quasi-spherical foliation \cite{Bartnik93}.  This null 
quasi-spherical (NQS) formulation is well-adapted to modelling 
spacetimes containing a single black hole, extending from the black 
hole to null infinity.

The purpose of this paper is to describe the numerical algorithms we 
have used in our implementation of the NQS Einstein equations, and to 
present results of some accuracy tests of the code.  Interactive 
access to the data sets described here, and many other simulations, is 
available online at 
\texttt{http://gular.canberra.edu.au/relativity.html}.  More detailed 
discussions of the physical and geometric significance of the results 
of the code will be presented elsewhere.

{}From the numerical programming viewpoint, 
the most significant features of the code are:
\begin{enumerate}
\item
        a characteristic coordinate $z\sim t-r$ (cf.~\cite{Stellmacher38}) 
        plays the role of ``time'', with the numerical evolution in the 
        direction of increasing $z$;
   
\item
        the numerical grid is based on spherical polar coordinates on the 
        $z$-level sets $\cN_{z}$;
    
\item 
        $S^{2}$ dependencies are handled by a combination of spectral 
        coefficients with respect to a basis of spin-weighted spherical 
        harmonics (for spins $0,1,2$, corresponding to scalar, vector and 
        tensor harmonics); values of the field on a uniform grid in the 
        spherical polar coordinates $(\vartheta,\varphi)$; and Fourier 
        coefficients of the values on the polar coordinate grid;
                        
\item  
        a non-uniform radial grid along the outgoing null hypersurfaces 
        $\cN_{z},$ which compactifies future null infinity $\scri^{+}$ and 
        is adjusted dynamically so radial grid points approximately follow 
        the inward null geodesics;

\item
        reformulations of the hypersurface (radial) Einstein equations 
        which enable the numerical modelling of the asymptotic expansions 
        of fields near null infinity;
   
\item  
   4th order Runge-Kutta time evolution;
        
\item
        the characteristic transport (hypersurface, radial) equations are 
        treated as a system of ordinary differential equations and 
        integrated using an 8th order Runge-Kutta method 
        \cite{DormandPrince81};
        
\item
        an 8th order convolution spline is used for interpolation, for 
        computing radial derivatives, to realign the fields with the 
        dynamically varying radial grid, and for suppressing high frequency 
        modes;

\item
        a first order elliptic system on $S^{2}$ is solved at each radius 
        and time step by the conjugate gradient method, accelerated by a 
        geometrically natural spectral preconditioner.
\end{enumerate}

A number of numerical consistency checks suggest that most quantities 
of interest which are calculated by the code (eg., the NSQ metric 
functions, the connection coefficients and the Weyl curvature spinors) 
have relative errors of about $0.001\%$, for simulations where the
gravitational waves carry no more than 20\% of of the total spacetime 
mass.  Of course, greater accuracy is found for more nearly linear, 
weak field, simulations.  The major limiting factor in determining the 
accuracy appears to be the spherical harmonic resolution, currently at 
$L\le15$.  Although we have implemented routines which extend this to 
$L\le31$, a full implementation is not possible due to limitations in 
our present hardware.

The algorithm was initially developed and tested on a 300MHz DEC Alpha 
with 512Mb memory, and presently runs on a 300MHz Sun Ultra 2 with 
784Mb, with a typical (L=15) run taking between 2 and 4 days.  
Preliminary descriptions of these results are given in 
\cite{Bartnik96,BartnikNorton98,Bartnik99b,Norton96,Norton98}.

The Cauchy and characteristic initial value problems differ 
significantly in the nature of their appropriate initial conditions.  
The Cauchy problem initial data consists of the initial 3-metric and 
extrinsic curvature \cite{SmarrYork78}, 
whereas the appropriate initial data for a 
characteristic initial surface is just the null metric.  In the NQS 
case the null metric is
\begin{equation}
        ds^{2}_{\cN_{z}} = (r\,d\vartheta+\beta^{1}\,dr)^{2} +
        (r\sin\vartheta\,d\varphi+\beta^{2}\,dr)^{2},
        \label{ds2Nz}
\end{equation}
parameterised by the angular shear vector $\beta = 
\frt(\beta^{1}-\imu\beta^{2}) = \beta(z,r,\vartheta\,\varphi)$.  
Unlike the Cauchy problem, the NQS Einstein equations 
\cite{Bartnik97a} do not impose any additional algebraic or 
differential constraints on $\beta$, so the initial data $\beta(z=0)$ 
is an arbitrary function of the spherical polar coordinates 
$(r,\vartheta\,\varphi)$, except for a mild size constraint 
\bref{bsize}.  Heuristically $\beta$ represents the in-going 
gravitational radiation of the spacetime; an interpretation which is 
consistent with the initial ($z=0$) values of $\beta$ being freely 
specifiable.  Note that the geometric invariant 
$\sigma_{NP}/\rho_{NP}$ \cite{Penrose72} becomes 
$-2\edth\beta/(2-\div\beta)$ in the NQS coordinates \cite{Bartnik97a}.

The NQS geometric gauge provides a formulation of the Einstein 
equations as an explicit \emph{characteristic transport} system, 
coupled to a time evolution equation.  This type of structure is also 
found in characteristic formulations of other hyperbolic equations 
such as the wave equation, and it is well-known for the Einstein 
equations in other characteristic coordinate based gauges 
\cite{Bondi62, Sachs62, Stellmacher38, NewmanUnti63}.  Although there are 
existence results for characteristic initial value formulations of 
hyperbolic equations, these rely on reduction to the Cauchy problem 
\cite{Rendall90, MullerzumHagen90} rather than directly on the transport 
form.  The only exception of which we are aware is the analysis of the 
linear wave equation in \cite{Balean96, Balean97}.  It would be 
valuable to have theoretical existence results for systems of 
characteristic transport equations, which could justify the numerical 
formulation described here.

The characteristic-based approach has recently been strongly advocated 
by Winicour and his coworkers 
\cite{Bishop92,GomezEt92,GomezWinicour92b}, who have developed codes 
for solving the scalar wave equation \cite{GomezEt92}, axially 
symmetric spacetimes \cite{GomezEt94a} and for the full Einstein 
equations \cite{BishopEt97b}, based on the Bondi coordinate system 
\cite{Bondi62}.  These works have been fundamental in establishing the 
feasibility of numerical formulations based on a characteristic 
coordinate and in motivating the present implementation.  However, the 
stability analyses and experience of \cite{GomezEt92,GomezEt94a} are 
not directly applicable to our evolution procedure, since the 
formulations and numerical methods used have several significant 
differences.  Like the Bondi parameterisation, the gauge conditions 
are directly implemented in the metric form; however, the NQS Einstein 
equations are considerably simpler than the Bondi coordinate form of 
the equations \cite{Bartnik97a,BishopEt97b}.

The treatment of angular derivatives is greatly simplified by the 
quasi-spherical condition, which encourages the use of spherical 
harmonic expansions.  These in turn will simplify comparisons with 
theoretical results on perturbations of the Schwarzschild black hole 
\cite{ReggeWheeler57,Zerilli70,Moncrief74,Chandrasekhar84}.  We note 
that the power of spectral methods in practical applications is 
well-known, in fields such as meteorology \cite{Swarztrauber89}, 
astrophysics \cite{BonazzolaEt98}, and fluid dynamics 
\cite{CanutoEt88}.  Finally, the combination of the 
characteristic-transport and method of lines techniques, considerably 
simplifies the use of high order algorithms such as RK4 for time 
evolution.

The method of lines approach to evolution equations is common in fluid 
dynamics \cite{CanutoEt88}, but has not previously been attempted in 
numerical relativity.  The situation we consider here, of smooth 
variations in a single black hole geometry, is well-suited to the 
method, since high frequency modes are less likely to be of physical 
interest and thus may be treated by smoothing or artificial viscosity.  

The use of higher order methods, together with spherical harmonics and 
radial smoothing, leads to considerably more accurate results than 
would be possible with the 2nd order methods more commonly employed.  
However, this suggests that our code is restricted to very smooth 
spacetimes, and cannot reliably treat spacetimes with strong localised 
features (such as planets, or gravitational shocks).  Because we are 
concerned primarily with the gravitational wave perturbation modes of 
a single black hole, this does not present an important restriction, 
since the dominant modes are known to occur only for low angular 
momentum $l$ -- in particular, the $l=2,3,4$ modes are expected to 
carry practically all the radiated gravitational energy.

The code models evolution in an exterior domain, and consequently 
boundary data must also be prescribed at an interior boundary surface 
$r=r_{0}$.  In general only two pieces of the boundary data can be 
freely specified, thereby fixing the null hypersurfaces $\cN_{z}$ and 
the outgoing radiation flux at the boundary \cite{Bartnik97a}.  The 
rest of the boundary data is constrained by boundary evolution 
equations (see \bref{eq:Gnn},\bref{eq:Gnm}).  For simplicity, the 
version of the code reported here assumes fixed boundary conditions, 
corresponding to the past horizon of a Schwarzschild black hole.  This 
assumption considerably simplifies the treatment of the inner 
boundary, and is completely consistent with the Einstein evolution.  
It follows that the code models the interaction of gravitational waves 
with a single Schwarzschild-like black hole.  Future versions of the 
code will incorporate dynamical inner boundary conditions.

The paper is organised as follows.  

Section 2 gives the NQS form of the equations, and describes the 
structure of the equations and the formal steps in the solution 
algorithm.  The geometric significance of the resulting equations is 
described in \cite{Bartnik97a}.

Section 3 describes the numerical techniques used, including the 
representation of spin-weighted spherical harmonics used to encode the 
angular variation of the various fields, and the high order 
convolution splines used for interpolation and differentiation in the 
radial direction.
 
Two aspects of the treatment of spherical fields appear to be 
non-standard \cite{Orszag74,Swarztrauber89}: the use of FFT's in both 
the $\varphi$ and $\vartheta$ directions, based on the ``torus'' model 
of $S^{2}$ \cite{Norton96}; and the use of \emph{spin-weighted} 
spherical harmonics to handle, in a unified and frame-invariant 
manner, vector and higher rank tensor harmonics as well as invariant 
derivative operators.

Section 4 describes the various stages in the evolution algorithm ---
solving the hypersurface equations out to null infinity $\scri^{+}$; 
reconstructing the metric from the connection variables (which 
includes the solution of a 1st order elliptic system on the 2-sphere 
at each radial grid position); and the evolution of the primary field 
$\beta$.

In section \ref{sec:accuracy} we describe various techniques for 
estimating the accuracy of the code, testing both numerical and 
geometric properties of the numerical solution.  Numerical convergence 
tests estimate the effects of separately increasing the resolution in 
the radial, time and angular directions.  The two geometric tests 
described here demonstrate the consistency of the numerical metric by 
verifying the constraint equations for the Einstein tensor components 
$G_{nn},G_{nm}$ (\ref{eq:Gnn},\ref{eq:Gnm}), and by testing the 
accuracy of the solution at infinity using the Trautman-Bondi mass 
decay formula \cite{Trautman58b,Trautman62,Bondi60,Bondi62}.  These 
provide highly nontrivial tests of the consistency of the numerical 
solution.

Other tests of the code, based on geometric properties of vacuum 
spacetimes, and comparisons with known exact and approximate solutions 
of the Einstein equations, are envisaged for future work.

\section{Einstein equations and NQS metric functions}
\label{sec2}

\subsection{Spacetime metric}

We consider spacetimes admitting global null-polar coordinates 
$(z,r,\vartheta,\varphi)$ in which the metric takes 
the null quasi-spherical form \cite{Bartnik97a}
\begin{equation}
        ds^{2}_{NQS} = -2u\,dz(dr+v\,dz) +
        (r\,d\vartheta+\beta^{1}\,dr+\gamma^{1}\,dz)^{2} +
        (r\sin\vartheta\,d\varphi+\beta^{2}\,dr+\gamma^{2}\,dz)^{2},
    \label{ds2:nqs}
\end{equation}
where $u>0,v$, and 
$\beta=\beta^{1}\partial_{\vartheta}+\beta^{2}\csc\vartheta\partial_{\varphi}$,
$\gamma=\gamma^{1}\partial_{\vartheta} 
  + \gamma^{2}\csc\vartheta\partial_{\varphi}$
are the unknown NQS metric functions, to be determined by numerical solution
of the Einstein equations. Note that we use $\partial_{\vartheta}, 
\partial_{\varphi}$ to denote equivalently the coordinate tangent 
vectors and the coordinate partial differential operators 
$\frac{\partial}{\partial \vartheta}$, $\frac{\partial}{\partial \varphi}$.
We may consider $\beta,\gamma$ either as vector fields on $S^{2}$ or as 
spin~1 quantities, defined by the complex combinations \cite{Goldberg67}
\begin{equation}
\label{vector}
    \beta = \frt (\beta^{1}-\imu\beta^{2}),\qquad
    \gamma = \frt (\gamma^{1}-\imu\gamma^{2}).
\end{equation}

The canonical example of a spacetime with metric in NQS form
is Schwarzschild spacetime
in Eddington-Finkelstein retarded coordinates \cite{dInverno92,HawkingEllis73}
\begin{equation}
        ds^{2}_{Schw} = - 2 dz\,(dr+\half(1-2\mschw/r)\,dz) + 
                   r^{2}(d\vartheta^{2}+\sin^{2}\vartheta\,d\varphi^{2})
\label{ds2:schw}
\end{equation}
with $u=1$, $v=\half(1-2\mschw/r)$, $\beta^{A}=\gamma^{A}=0$ and 
$\mschw=const$.  This includes Minkowski space $\bR^{3,1}$ as the case
$\mschw=0$ and $z=t-|x|$.

\subsection{Edth}

Using the complex notation \bref{vector}, we have the canonical 
angular covariant derivative operator ``edth'' 
\cite{Goldberg67,PenroseRindler84,EastwoodTod82},
\begin{equation}\label{edth}
        \edth \eta = \frac{1}{\sqrt{2}}\sin^s \vartheta \left( \frac{\partial }
        {\partial \vartheta} - \frac{\imu}{\sin \vartheta} 
        \frac{\partial }{\partial \varphi}
        \right)\left(\eta\,\sin^{-s}\vartheta\right)
        =
        \frac{1}{\sqrt{2}}\left( \frac{\partial }{\partial \vartheta} 
          - s\cot\vartheta
          - \frac{\imu}{\sin \vartheta} \frac{\partial }{\partial \varphi}
        \right)\,\eta ,
\end{equation}
acting on a spin $s$ field $\eta$, and its ``conjugate'' operator 
$\edthbar$,
        \begin{equation}\label{edthbar}
        \edthbar \eta 
        =\frac{1}{\sqrt{2}}\left( \frac{\partial }{\partial \vartheta} 
          + s\cot\vartheta
          + \frac{\imu}{\sin \vartheta} \frac{\partial }{\partial \varphi}
\right)\,\eta .
\end{equation}
All geometric angular derivative operators may be defined in terms of
$\edth$, $\edthbar$.  For example, the covariant directional derivative
of a spin $s$ field $\eta$ in the direction $\beta$ is
\[
\nabla_{\beta}\eta = \beta\edthbar \eta + \bar{\beta}\edth \eta\ ;
\]
the divergence and curl (of a vector) are
\begin{eqnarray}
        \div\beta & = & \edthbar\beta + \edth\bar{\beta} = 
        \nabla_{1}\beta^{1}+\nabla_{2}\beta^{2},
        \label{div}  \\
        \curl \beta & = & \imu (\edthbar\beta-\edth\bar{\beta}) = 
        \nabla_{2}\beta^{1}-\nabla_{1}\beta^{2} \ ;
        \label{curl}
\end{eqnarray}
and the spherical Laplacian is
\begin{equation}\label{laplace}
   \Delta \eta = (\edth\edthbar + \edthbar\edth)\eta \ .
\end{equation}
Further properties of edth are described in section \ref{sec3} and in 
\cite{EastwoodTod82,PenroseRindler84,Bartnik97a}.
\subsection{Connection variables}
\label{sec2.3}

In addition to the metric functions $(u,v,\beta,\gamma)$ we introduce the 
connection fields $H,J,K,Q,\Qpm$
\begin{eqnarray}
H  &=& \frac{1}{u}(2- \div\beta), \label{def:H} \\[5pt]
J  &=& v(2- \div\beta) + \div \gamma, \label{def:J} \\[5pt]
K  &=& v\edth \beta - \edth \gamma, \label{def:K} \\[5pt]
Q  &=& r\frac{\partial \beta}{\partial z} - r \frac{\partial \gamma}{\partial r}
     + \gamma + \nabla_\beta \gamma-\nabla_{\gamma}\beta, \label{def:Q}\\[5pt]
\Qpm &=& \frac{1}{u}(Q \pm \edth u). \label{def:Qpm} 
\end{eqnarray}
Observe that $u,v,H,J$ are real and have spin~0, whereas 
$\beta,\gamma,Q,\Qp,\Qm$ have 
spin~1 and $K$ has spin~2.  


Given the metric functions $u,v,\beta,\gamma$ on a $z$-level set $\cN_{z}$, 
we may construct $H,J,K$ on $\cN_z$ directly, 
and $Q$ (and $\Qpm$) may be reconstructed if in addition,
$\partial\beta/\partial z$
is also known on $\cN_{z}$.  It is clear from 
(\ref{def:H}--\ref{def:Qpm}) that this construction does not 
require any compatibility conditions on the data 
$u,v,\beta,\gamma,\frac{\partial}{\partial z}\beta$.  

Rather remarkably, there is a converse construction for the metric 
functions $u,v,\gamma$ and $\partial \beta/\partial z$, which also 
involves totally free and unconstrained data, namely the connection 
variables $H,J,K,Q$.  This contrasts sharply with the description of 
the connection via the Newman-Penrose spin coefficients 
\cite{NP62,PenroseRindler84}, which requires numerous differential 
constraint equations, expressing the property that the connection is 
{}torsion-free.

The converse construction works as follows.  Given $\beta$ and the 
connection variables ($H,J,K$) on $\cN_{z}$, we reconstruct $u$ via 
the relation
\begin{equation}
        u = \frac{2-\div\beta}{H},
\label{u:def}
\end{equation}
and we find $v,\gamma$ by solving an elliptic system for $\gamma$,
\begin{equation}
    \cL_{\beta}\gamma := 
    \eth \gamma + \frac{\eth \beta}{2 - \div \beta}\,\div \gamma 
    = J  \frac{\eth \beta}{2 - \div \beta} - K,
\label{Lbeta:def}
\end{equation}
and setting
\begin{equation}
        v = \frac{J-\div\gamma}{2-\div\beta}.
\label{v:def}
\end{equation}
The system (\ref{Lbeta:def}) is $\bR$-linear and elliptic with 
6-dimensional kernel, provided $\edth\beta$ is not too large 
($|\edth\beta| < (2-\div\beta)/\sqrt{3}$ is sufficient).  Prescribing 
the $l=1$ spherical harmonic coefficients of $\gamma$ (for example, by 
requiring $\gamma_{l=1}=0$) suffices to determine the solution 
$\gamma$ uniquely.  The remaining connection parameter $Q$, together 
with the now known values of $\beta,\gamma$ on $\cN_{z}$, determines 
the \emph{evolution equation}
\begin{equation}
\label{dbdz:eq}
                \frac{\partial \beta}{ \partial z } =
    \frac{\partial \gamma}{ \partial r }
    + \frac{1}{r}( Q+ \nabla_\gamma \beta -\nabla_\beta \gamma -\gamma).
\end{equation}

{}To summarise, given the field $\beta$ on a single level 
set $\cN_{z}$, satisfying the size constraint
\begin{equation}
|\edth\beta| < (2-\div\beta)/\sqrt{3},
\label{bsize}
\end{equation} 
the map $(u,v,\gamma,\beta_{z})\mapsto (H,J,K,Q,\gamma_{l=1})$ 
is invertible, assuming all fields are sufficiently smooth.  In 
section \ref{sec:elliptic} we will describe the numerical implementation of 
the inverse map.

\subsection{NQS Einstein equations}
{}To compute the curvature of $ds^{2}_{NQS}$, and thereby to determine the 
NQS form of the Einstein equations,  we introduce the complex null
vector frame
$(\ell,n,m,\mbar)$,
\begin{eqnarray}
        \ell &=& \partial_{r}-r^{-1}\beta,
\nonumber\\
        n &=& u^{-1}(\partial_{z}-r^{-1}\gamma
                -v(\partial_{r}-r^{-1}\beta)),
\label{tetrad:def}
\\
        m&=&\frac{1}{r\sqrt{2}}
        (\partial_{\vartheta}-\imu\csc\vartheta\,\partial_{\varphi}),
\nonumber
\end{eqnarray}
and the directional derivative operators 
\begin{equation}
\cD_r = \partial_{r}- r^{-1}\nabla_{\beta}\,,
\quad \quad  \cD_z = \partial_{z}- r^{-1}\nabla_{\gamma}.
\label{DrDz}
\end{equation}
Expressions for the Newman-Penrose spin coefficients
\cite{NP62} with respect to the frame $(\ell,n,m,\mbar)$
are given in terms of $H,J,K,Q$ in \cite{Bartnik97a}.

The frame components of the Einstein tensor $G_{ab}$, 
$a,b=\ell,n,m,\bar{m}$,
may be written in 
terms of the NQS metric functions 
and NQS connection variables. 
These expressions may be grouped into 
{\em hypersurface equations} (or \emph{main equations} 
\cite{Bondi62,Sachs62a}):
\begin{eqnarray}
r \cD_{r}H&=& 
   \left(\half\div\beta -
       \frac{2|\edth\beta|^2+r^{2}G_{\ell\ell}}{2-\div\beta} \right)H,
\label{eq:Gll}
\\[3pt]
r\cD_{r}\Qm &=& 
  (\edth\bar{\beta}-uH)\Qm + \Qmbar\edth\beta + 2\edthbar\edth\beta
{}+ u\edth H - H\edth u  + 2r^{2} G_{\ell m} ,
\label{eq:Glm} 
\\[3pt]
r \cD_{r}J &=& 
   -( 1-\div\beta)J + u - \half u |\Qp |^2
     {}-\half u\,\div(\Qp ) - u r^2 G_{\ell n} ,
\label{eq:Gln} 
\\[3pt]
r \cD_{r}K &=& 
\left(\half \div\beta + \imu\curl\beta \right) K -\half J \edth\beta 
{}+ \half u \edth \Qp  +\quarter u (\Qp )^2 +\half ur^2G_{mm},
\label{eq:Gmm}
\end{eqnarray}
the {\em boundary equations} (or \emph{subsidiary equations})
\begin{eqnarray}
r\,\cD_z\left({J/u}\right) &=& 
  {v^2}\,r\cD_r\left(J/(uv)\right)
 +\half(\div\gamma - v\,\div\beta)J/u  +   2u^{-1}|K|^{2}
 \nonumber\\ && {}
   - \nabla_{Q^{+}}v - \Delta v + u {r^2} G_{nn} , 
\label{eq:Gnn}
\\[3pt]
r\,\cD_z \Qp &=&
 (v\,r\cD_r + J 
-v\edth\bar{\beta}+\edth\bar{\gamma})\Qp
   - K\bar{Q}^{+}
   + 2\edthbar K + 2u^{-1}r\cD_{r}(u\edth v)
\nonumber\\ && {}   
   -(2 + \imu \curl\beta)\edth v
  +\edth J 
  -2u^{-1}J\edth u
  - 2u r^{2} G_{nm} ,
\label{eq:Gnm}
\end{eqnarray}
and the {\em trivial equation}
\begin{eqnarray}
\label{eq:Gmb}
        u r^{2}G_{m\bar{m}} & = & 
        r\cD_{r}J -\half \div \beta\,J - u|Q^{+}|^{2} +\half u\,\div Q^{+} 
           +      \bar{Q}^{+}\edth u + Q^{+}\edthbar u  
\nonumber 
\\[1pt]
&  & {}
     + \bar{K}\edth\beta + K \edthbar\bar{\beta}
         + r^{2}(v \cD_{r} - \cD_{z})(u^{-1}\cD_{r}u)  
                + u^{-1}r^{2}\cD_{r}(u\cD_{r}v).
\end{eqnarray}


Observe that the hypersurface equations (\ref{eq:Gll}--\ref{eq:Gmm})
have no explicit $z$-derivatives, and they each contain only one
radial ($r$) derivative.  The form of the connection variables
($H,J,K,Q$) was determined by exactly these properties.  Consequently,
the hypersurface equations may be written schematically in terms of
$U=(H,\Qm,J,K)$ in the ``characteristic-transport'' form
\begin{equation}
   r\frac{\partial U}{\partial r} = F(\beta(z,r),U(z,r)) ,
\label{transport}
\end{equation}
by treating angular derivatives such as $\edth U$ as determined by the
set of values $U(z,r)$ on the full $S^{2}$. This system has the
effect of transporting the fields $U$ along the characteristic curves
with tangent vector $\ell$ which foliate the null hypersurfaces
$\cN_{z}$.

Note that alternative reformulations of the hypersurface equations are 
possible, preserving the general characteristic transport structure.  
For reasons associated with reliably capturing  the asymptotic behaviour of 
the fields, the present version of the code integrates the following radial 
equations, for the variables $\log (H/2)$ (instead of $H$), 
$j=\tfrac{1}{4} r(1-HJ)-\mschw$ (instead of $J$) and $r\Qp$ (instead of $\Qm$):
\begin{eqnarray}
    r\partial_{r}\log (H/2) & = & \nabla_{\beta}\log (H/2) +
     \half\div\beta - \frac{2|\edth\beta|^2+r^{2}G_{\ell\ell}}{2-\div\beta} ,
\label{eq:lh}
\\
    r\partial_{r}(r\Qp) & = & \nabla_{\beta}(r\Qp) -
        (1-2\edth\bar{\beta})(r\Qp) +\nabla_{r\Qp}\beta 
        +2r\edth(r\cD_r\log u)
\nonumber
\\
&&{}+2r\beta-\imu r\edth\curl\beta +2r^{3}G_{\ell m} ,
\label{eq:rQp}
\\
        r\partial_{r}j & = & \nabla_{\beta}j +
         (j+\mschw-\half r)(\div\beta - r\cD_r\log u)
\nonumber
\\
&&{}     +\tfrac{1}{4} r (|\Qp|^{2}+\div\Qp) + \half r^{3} G_{\ell n} ,\ 
\label{eq:j}
\\
        2r\partial_{r}K & = & 2\nabla_{\beta}K +
        \left(\div\beta + 2\imu\curl\beta \right) K - J \edth\beta 
\nonumber
\\
&&{}     + u \edth \Qp  +\half u (\Qp )^2 + ur^2G_{mm}.
\label{eq:K}
\end{eqnarray}
Here $\mschw=1$ is a constant which fixes the bare mass of the 
background Schwarzschild black hole.  Of course, the Einstein tensor 
components in these formulae are set to zero for the vacuum equations.


It is remarkable that the boundary equations 
(\ref{eq:Gnn},\ref{eq:Gnm}) and the trivial equation \bref{eq:Gmb} may 
be regarded as \emph{compatibility} relations, by virtue of the 
conservation (contracted Bianchi) identity $G_{ab}^{\ \ ;b}=0$ 
\cite{Sachs62a,Bartnik97a}.  This identity is valid for any Einstein tensor 
$G_{ab}$, regardless of the metric.  Substituting the hypersurface 
equations $G_{\ell\ell}=G_{\ell m}=G_{\ell n}=G_{mm}=0$ into the 
conservation identity, yields equations $HG_{m\mbar}=0$ and a 
propagation system for $G_{nn},G_{nm}$ which has the unique solution 
$G_{nn}=G_{nm}=0$ if the boundary equations are satisfied on one 
hypersurface transverse to the outgoing null surfaces $\cN_{z}$.  Thus 
in order to construct a solution of the full vacuum Einstein 
equations, it suffices to satisfy the hypersurface equations 
everywhere, and the boundary equations just on the boundary surface 
$r=r_{0}$ (for example).

\section{Numerical techniques}
\label{sec3}

In this section we describe the data representation and manipulation 
techniques.  These consist mainly of techniques for handling angular 
fields and derivatives, and an unusual convolution spline used for 
interpolation, differentiation and high frequency filtering in the 
radial and time directions.

\subsection{Fields on $S^{2}$}

The evolution algorithm treats the angular derivatives $ 
\frac{\partial }{ \partial \vartheta}, \frac{\partial }{ \partial 
\varphi}$ as ``lower order'', compared to the radial and time 
derivatives $ \frac{\partial }{ \partial r}, \frac{\partial }{ 
\partial z}$.  This attitude in a numerical computation can be 
justified only if it is possible to easily and accurately compute and 
manipulate angular derivatives.  This is achieved by using spectral 
representations (both Fourier and spherical harmonic) for fields on 
the 2-sphere.  This approach is widely used in geophysical and 
meteorology applications 
\cite{Merilees73b,Orszag74,Boyd78a,Swarztrauber79,Swarztrauber96} and 
is known to have significant advantages compared to finite difference 
approaches \cite{Swarztrauber89}, based on either angular coordinate 
grids or overlapping stereographic projection charts 
\cite{Starius77,Singleton90,GomezEt97a}.  Nevertheless, spectral 
methods have rarely been used in numerical general relativity 
(however, see \cite{Prager96,BonazzolaEt98}) and they have not been 
used previously for solving the full Einstein equations.

The basic manipulations required of $S^{2}$ fields are:
\begin{itemize}
\item  computing non-linear algebraic terms such as $1/(2-\div\beta)$, 
$u|\Qp|^{2}$ etc; 
\item  computing angular derivatives operators 
such as $\div\beta,\ \edth\Qp$ etc; 
\item inverting the linear elliptic operator $\cL_{\beta}$ (which appears in 
 equation (\ref{Lbeta:def})); and
\item projecting aliased or noisy field value data onto certain subspaces 
of spin-weighted spherical harmonics.
\end{itemize}

{}To carry out these manipulations, three separate representations are
used for fields on $S^2$:
\begin{itemize}
        \item \emph{field values} $\eta_{jk}=\eta(\vartheta_{j},\varphi_{k})$
        at the polar coordinate grid points
\begin{equation}
        (\vartheta_{j},\varphi_{k})=((j-\half)\Delta\vartheta,(k-1)\Delta\varphi),
\label{phik}
\end{equation}
        where $\Delta\vartheta=2\pi/N$, $\Delta\varphi=2\pi/N$
        with $1\le j\le N/2,1\le k\le N$
        and (in our implementations) $N=16,\,32$ or $64$;

        \item \emph{Fourier coefficients} $\hat{\eta}_{mn}$ 
        arising from FFT transforms in the $\vartheta$ or $\varphi$
        directions, of the  field values $\eta_{jk}$;

        \item 
        spin-weighted  \emph{spherical harmonic coefficients} $\eta^{lm}$,
        $|m|\le l$, $l=s,\ldots,L$, $L=N/2-1$ (for spin $s=0,1$ and 2).
\end{itemize}

The field value representation is used when computing non-linear algebraic 
terms such as $u|\Qp|^{2}$.  The Fourier representation is used for 
computing $\vartheta$ and $\varphi$ angular derivatives, which are needed 
in the formulas for $\edth$, $\div$, for example. The spherical harmonic 
representation is used in solving the elliptic system (\ref{Lbeta:def}), 
{}to spectrally limit the field values by projection to spherical 
harmonic data, and to summarise the computation results (which are 
stored using spherical harmonic coefficients).

For fields which do not alias on the $(\vartheta,\varphi)$-grid, the 
three representations are completely equivalent in the sense 
that conversion between them is essentially exact, depending on the
machine precision and on algebraic details of the specific FFT 
algorithm used.  The requirement that a field does not alias is 
satisfied when it can be represented by a finite expansion in 
spin-weighted spherical harmonics with angular momentum $l\le 
L=N/2-1$.  Our implementations use spectral cutoffs 
$L=15$ (for $N=32$) and $L=31$ (for $N=64$).

Transformation to the spherical harmonic representation involves a 
projection, because both the field value and Fourier representations 
have approximately twice as many degrees of freedom.  For example, a 
(real) spin~0 field with $l \le L= N/2-1$ has $(L+1)^2 = N^2/4$ 
spherical harmonic coefficients, whereas it has $N^2/2$ values on the 
$(\vartheta,\varphi)$-grid.  The space of non-aliasing spherical 
harmonics is a {\em linear} subspace of the space of functions 
represented by either Fourier coefficients or field values. 

For example, when the $(\vartheta,\varphi)$-grid field values of the 
product of two fields is calculated, the result, which will contain 
components up to $l \le 2L$, is aliased onto the grid in such a way 
that its field values no longer lie in the appropriate spin weighted 
spherical harmonic subspace.  To clean up after such non-linear 
effects, we project the result back onto the correct subspace, as 
described in section \ref{sph:sec}.

{}To minimise the possibility of unstable feedback 
of quadratic aliasing errors, we may invoke the Orszag 2/3 rule 
\cite{CanutoEt88, Orszag71} at various points within the code.  The 
effective spectral resolution of the code is then $l_{\rm max} \approx 2L/3$ 
($l_{\rm max} = 10$ for $N=32$ and $l_{\rm max} = 20$ for $N=64$).

\subsection{Spherical harmonics}

We first summarise the more important properties of $\eth$ 
(``edth'') and spin weighted spherical harmonics.  The edth formalism 
provides a unified geometric approach to the treatment of angular 
derivatives on $S^{2}$ and vector and higher-rank tensor harmonics.  
Detailed descriptions of the properties of spin-weighted 
fields and spherical harmonics may be found in 
Penrose and Rindler \cite{PenroseRindler84} or 
\cite{Goldberg67,EastwoodTod82}.  Here we describe only the basic formulae.

We use a real-valued basis $Y_{lm}$, $l = 0,1,2,\ldots$, $m = -l,\ldots,l$ 
for the space of spin~0 spherical harmonic functions, defined by
\begin{equation}\label{Ylm}
            Y_{lm} = \overline{P}_{lm}(\vartheta)F_m(\varphi)\,,
\end{equation}
where
\begin{equation}\label{Fm}
                  F_m(\varphi) = \left\{ \begin{array}{cr}
                       1               & \quad m = 0 \\
                       \sqrt{2}\cos m\varphi & m > 0 \\
                       \sqrt{2}\sin |m|\varphi & m < 0,
\end{array}\right. 
\end{equation}             
and the $\overline{P}_{lm}(\vartheta) = \overline{P}_{l|m|}(\vartheta)$
are related to the associated Legendre functions $P_{lm}$ by
\begin{eqnarray}
\label{Pblm}
  \overline{P}_{lm}(\vartheta) &=& 
         (-1)^m \sqrt{2l+1} \sqrt{\frac{(l - m)!}{(l + m)!} }
        P_{lm}(\cos \vartheta)\,,          
\\
\label{Plm}
          P_{lm}(\cos \vartheta) &=& \frac{(-1)^m}{2^l l !}
 \sin^m\vartheta \left[ \frac{d^{l+m}}{dx^{l+m}}(x^2-1)^l
\right]_{x=\cos\vartheta}\,.    
\end{eqnarray}
The spin $s$ spherical harmonics $Y^{s}_{lm}$ are then defined explicitly by
\begin{eqnarray}\label{Yslm}
        Y^{s}_{lm}& = & \phantom{(-1)^{s}} 
            \left[\frac{2^{s}(l-s)!}{(l+s)!}\right]^{1/2}
             \edth^{s}Y_{lm}\,,\qquad s>0,
        \label{Y+slm}  \\
        Y^{-s}_{lm}& = & (-1)^{s} \left[\frac{2^{s}(l-s)!}{(l+s)!}\right]^{1/2}
             \edthbar^{s}Y_{lm}\,,\qquad -s<0,
        \label{Y-slm}
\end{eqnarray}
where necessarily $l\ge |s|$.  
Note that the differential operator $\edth$ is spin-raising, sending spin $s$ into 
spin $(s+1)$ fields, and $\edthbar$ is spin-lowering,
\begin{eqnarray}
        \edth Y^{s}_{lm} & = & \left[\half(l+s+1)(l-s)\right]^{1/2} 
        Y^{s+1}_{lm},
        \label{edthYslm}  \\
        \edthbar Y^{s}_{lm} & = & -\left[\half(l+s)(l-s+1)\right]^{1/2} 
        Y^{s-1}_{lm},
        \label{edthbarYslm}
\end{eqnarray}
for all $s\in\bZ$, and $Y^s_{lm}$ and $Y^{-s}_{lm}$ are
related by complex conjugation,
\begin{equation}
\label{Y-s}
   Y^{-s}_{lm} = (-1)^{s} \bar{Y}^{s}_{lm}.
\end{equation}
Since $\Delta Y_{lm}= -l(l+1) Y_{lm}$, the fundamental commutation
relation 
\begin{equation}
        [\edthbar,\edth]\eta = (\edthbar\edth - \edth\edthbar)\eta = s \eta,
        \label{commutator}
\end{equation}
for any spin $s$ field $\eta$, may be used to show that
\begin{equation}
        \Delta Y^{s}_{lm} = (s^{2}-l(l+1)) Y^{s}_{lm},
        \label{lap-s}
\end{equation}
where $\Delta=\edth\edthbar+\edthbar\edth$.
With these conventions we have the orthogonality relations
\[
\frac{1}{4\pi}\int_{S^2}  Y^{s}_{lm}\, 
           \bar{Y}^{s}_{l'm'} \sin\vartheta\, d\vartheta d\varphi
=  \delta_{ll'} \delta_{mm'},
\]
which show that the $Y^{s}_{lm}$ form a basis (over $\bC$) of the 
Hilbert space of square-integrable spin $s$ fields on $S^{2}$, which 
is orthonormal in the natural Hermitian inner product
\begin{equation}
        \langle \phi,\psi\rangle = \frac{1}{4\pi}\oint_{S^{2}}\Re( \bar{\phi}\psi).
\label{s2ip}
\end{equation}

{}From (\ref{Ylm}--\ref{Plm}) it is evident that the spin~0 harmonics 
$Y_{lm}$ are trigonometric polynomials in $\vartheta$ and $\varphi$ 
\cite{Merilees73b}.  Using expression (\ref{edth}) for $\edth$, we also see 
that the $ Y^{s}_{l m}$ are trigonometric polynomials.   
The highest wave number Fourier modes which occur in the set of basis 
functions $\{Y^{s}_{l m} : s\leq l \leq L,\ |m|\le l \}$ are 
$\cos (L\vartheta),\ \sin (L\vartheta),\ 
\cos (L\varphi)$, and $\sin (L\varphi)$. Therefore, on a uniform 
$(\vartheta,\varphi)$-grid
of size $N/2\times N,$ one can represent all of the spin weighted
spherical harmonic functions up to $L = N/2 - 1$.

\subsection{Even/Odd decomposition}

Because $\edth$ is surjective onto the space of smooth spin $s$
fields for $s\ge1$ \cite{PenroseRindler84}, the decomposition of spin~0
fields or functions into real and imaginary parts may be propagated to
higher spin.  The resulting decomposition into \emph{even} and
\emph{odd} components plays an important role in the analysis of the
linearised Einstein equations about the Schwarzschild spacetime
\cite{ReggeWheeler57}.  Because the NQS geometry distinguishes the
Schwarzschild metric and is also based on spherical harmonics, it is
ideally suited to comparing nonlinear evolution to the comparatively
well-understood black hole linearised Einstein equations 
\cite{Chandrasekhar84, FuttermanEt88}.  It is thus
not surprising that the even-odd decomposition proves to be
very important in analysing the results of the NQS evolution.
 
We say that the spin $s$ field $\eta$ is 
\emph{even} if $\eta =\edth^{s}f$ for some real-valued function $f$, and 
$\eta$ is \emph{odd} if $\eta =\imu\edth^{s}g$ for some real-valued
function $g$.  (If $s<0$ then we interpret $\edth^s$ as
$(-\edthbar)^{|s|}$).  This matches the usage in \cite{ReggeWheeler57}
--- note that for axially symmetric fields the terms
\emph{polar} (for even) and \emph{axial} (for odd) are sometimes used
\cite{Chandrasekhar84}.  The surjectivity of $\edth$ onto spin $s\ge1$
ensures that every spin $s$ field may be uniquely decomposed into a
sum of even and odd parts.  For $s=1$ this decomposition corresponds
exactly to the classical Hodge-Helmholtz decomposition of a vector
field into the sum of a gradient  and a dual gradient (or curl) --- see Table 
\ref{table:odd-even} for a summary of the various nomenclatures.
\begin{table}[ht]
        \centering
\begin{tabular}{c|c|c|c|c|c}
        Even: & polar & irrotational & $\edth f$ & $\textrm{grad}f$ & 
        $(\nabla_{1}f)\,v_{1}+(\nabla_2f)v_2 $ \\
        \hline
        \hline
        Odd: & axial & divergence-free & 
        $\imu \edth g$ & $\curl g$ & 
        $(\nabla_{2}g)\,v_{1}-(\nabla_{1}g)v_{2}$  \\
\end{tabular}
        \caption{Equivalent terminologies for vector fields on $S^{2}$} 
        \label{table:odd-even}
\end{table}

The even/odd decomposition has a natural interpretation in terms of
the spectral decomposition
\begin{equation}
\label{eq:shc}
    \eta = \sum_{l=s}^{\infty} \sum_{m=-l}^{l} \eta^{lm} Y^{s}_{lm} 
\end{equation}
of a spin $s$ field $\eta$, because we are using a basis of real-valued 
$Y_{lm}$.  Namely, $\eta$ is \emph{even} if the spectral coefficients 
$\eta^{lm}$ are real, and \emph{odd} if the coefficients are pure 
imaginary.  We sometimes use $\textrm{Even}(\eta)$ and $\textrm{Odd}(\eta)$ 
{}to represent the respective projections, so $\eta = \textrm{Even}(\eta) + 
\textrm{Odd}(\eta)$ with
\begin{eqnarray}
\label{even}
        \textrm{Even}(\eta) & = & \sum\Re(\eta^{lm}) Y^{s}_{lm}
\\
\label{odd}
        \textrm{Odd}(\eta) & = & \imu \sum\Im(\eta^{lm}) Y^{s}_{lm}.
\end{eqnarray}

Observe that $\edth Y^{s}_{sm}=0$ for $s\ge0$ and thus $\edth$ acting on 
spin $s$ fields with $s\ge0$ has kernel having (complex) dimension $2s+1$.  
Likewise the formal adjoint $-\edthbar$ acting on spin $s\le0$ fields has 
$(2|s|+1)$-dimensional kernel.  In particular, $\edth$ acting on spin~1 
fields has kernel consisting of the $\bC$-linear space spanned by the three 
$l=1$ spin~1 spherical harmonics $Y^{1}_{1m}$ --- the corresponding real 
vector fields are the dual gradients of functions linear in $\bR^{3}$, 
which are just the infinitesimal rotations, and the gradients, which are 
the conformal dilation vector fields.

The correspondence between vector fields on $S^2$ and spin~1 fields 
generalises to spin~2 fields, which correspond to symmetric traceless 
2-tensors on $S^{2}$.  If $\lambda$ is a symmetric traceless 2-tensor 
then with respect to the standard polar coordinate derived orthonormal 
frame
\begin{equation}
        e_{1} = \partial_{\vartheta},\quad e_{2} = 
        \csc\vartheta\,\partial_{\varphi},
\label{e12}
\end{equation}
we have the correspondence
\begin{equation}
        \lambda \sim \half (\lambda^{11}-\lambda^{22} - 
        2\imu\lambda^{12})\,.
        \label{2tensor}
\end{equation}
This correspondence extends to higher integer spins with higher rank 
symmetric traceless tensors on $S^{2}$.  The cases $s=0,1,2$ 
of most importance in our work correspond to the usual scalar, 
vector and tensor harmonics.

\subsection{Fourier representation}

The fast Fourier transform (FFT) is used to transform between Fourier 
coefficients and field values on the uniform 
$(\vartheta,\varphi)$-grid.  Fourier convergence problems arising from 
discontinuities in coordinate derivatives and vector and tensor 
components at the poles, are sidestepped by an observation relating 
fields on $S^{2}$ to fields on the torus ${\bT}^{2}=S^{1}\times S^{1}$ 
\cite{Norton96}.  The torus method enables coordinate and covariant 
derivatives for all types of field to be computed using Fourier 
methods, so is particularly well-suited to handling the derivative 
operator $\eth$ \bref{edth}.  This approach to handling component 
discontinuities at the poles is simpler than the techniques reviewed 
in \cite{Swarztrauber81} for manipulating vector fields, and readily 
extends to any rank $s\ge 0$.
    
For integer $s$ the real and imaginary parts of a spin $s$ field on 
$S^2$ may be identified with the two independent frame components of a 
completely symmetric trace-free tensor of rank $|s|$ on $S^2$ 
\cite{PenroseRindler84}.  Since the frame $e_{1},e_{2}$ (\ref{e12}) is 
not continuous at the poles, the tensor components will not be 
continuous at the poles, so are not obviously suited to Fourier 
expansion in the $\vartheta$ direction.

However, along any smooth curve crossing through a pole, both basis vectors
$e_1$ and $e_2$ reverse direction at the pole.  Thus, for any smooth tensor 
field
$T = T^{j_1 \ldots j_s} e_{j_1}\otimes\cdots\otimes e_{j_s}$,
by continuity of $T$ the component functions $T^{j_1 \ldots j_s}$
will change by a factor $(-1)^{s}$ across the poles.  Consequently, if we
extend the domain of definition of $T^{j_1 \ldots j_s}$ to $\vartheta\in 
[-\pi,\,\pi]$ by
\begin{equation}\label{T-torus}
    T^{j_1 \ldots j_s}(-\vartheta,\varphi) \;=\; (-1)^s T^{j_1 \ldots
     j_s}(\vartheta,\varphi+\pi)\,, \quad \mathrm{for }\ \vartheta \in
     [0,\pi]
\end{equation}
(using the $2\pi$ periodicity in $\varphi$), then the resulting
extension is $2\pi$-periodic and continuous in $\vartheta$.  This
argument extends to higher (covariant) derivatives of $T$, showing
that the extension is in fact \emph{smooth} and periodic in
$\vartheta$.  Derivatives of $T^{j_1 \ldots j_s}$ with respect to
$\vartheta$ can then be calculated just as for $\varphi$ derivatives,
provided that the direction of increasing $\vartheta$ is 
properly taken into account.

In effect, the extension just described defines a (smooth) field 
$T^{j_1\ldots j_s}$ on the torus $\bT^{2}=S^1 \times S^1$.  This may be 
understood geometrically by noting that the map
\begin{equation}
   \Upsilon:{\bT}^{2}\to S^{2}, \quad (\vartheta,\varphi)\mapsto 
   \left\{ 
      \begin{array}{ll}
         (\vartheta,\varphi),    &\vartheta\in (0,\pi], \ \phi\in (-\pi,\pi] \\
         (-\vartheta,-\varphi),  &\vartheta\in (-\pi,0],\ \phi\in (-\pi,\pi] \\
      \end{array}\right.
\label{torusmap}
\end{equation}
is in fact \emph{smooth}.  This follows by noting that because 
$\vartheta$ is a radial coordinate near the north pole $\vartheta=0$, 
the differential structure near the pole is represented by the 
rectangular coordinates $(\xi,\eta) = 
(\vartheta\cos\varphi,\vartheta\sin\varphi)$ and the map 
$(\vartheta,\varphi)\mapsto (\xi,\eta)$ is manifestly $C^{\infty}$ for 
$\vartheta$ near $0$.  Consequently any smooth tensor $T$ on $S^{2}$, 
when expressed in a cotangent basis, pulls back to a smooth tensor on 
${\bT}^{2}$ (ie.~$\Upsilon^{*}(T)\in C^{\infty}({\bT}^{2})$), and thus 
admits a well-behaved Fourier representation on ${\bT}^{2}$.  The 
converse is of course false: a smooth tensor on ${\bT}^{2}$ does not 
necessarily arise from a smooth tensor on $S^{2}$, even if it 
satisfies the parity condition \bref{T-torus} satisfied by pull-back 
tensors.

The coordinate derivative form \bref{edth} of $\edth$ (and similarly any 
other covariant angular derivative operator) can be easily 
evaluated by transforming to the Fourier representation of the field,
multiplying the Fourier coefficients by the appropriate wavenumber factors, 
transforming back to obtain the field
value representation of the $\vartheta$ and $\varphi$ derivatives,
then finally including the $\csc\vartheta$ factors.  Thus, computing a 
derivative 
operator such as $\edth$ is an $O(N^{2}\log N)$ operation.
Since 
$\csc(\half\Delta\vartheta)\simeq 10$ for $N=32$, there is no 
significant loss of accuracy in calculations near the poles.

It should be emphasised that because the $Y^{s}_{l m}$ are 
trigonometric polynomials in $(\vartheta,\varphi)$, the FFT 
computation of their numerical derivatives is {\em algebraically 
exact}.  For example, the Laplacian relation \bref{lap-s} is 
numerically verified to 1 part in $10^{12}$ \cite{Norton96}.
  
\subsection{The spherical harmonic representation}
\label{sph:sec}

The field value and Fourier representations suffice for most numerical 
calculations, such as computing the nonlinear terms in 
(\ref{eq:Gll}--\ref{eq:Gmm}).  However, at some steps it is essential 
{}to use the representation by spherical harmonic coefficients 
\bref{eq:shc}:
\begin{enumerate}
\item 
     to solve the elliptic system \bref{Lbeta:def} (see section 
     \ref{sec:elliptic}); 
\item
     to implement spectral projections after each time step, with the aim 
     of suppressing aliasing effects and rounding errors at 
     the poles \cite{Orszag74};
\item 
     to store results for later analysis and display, since the 
     spherical harmonic representation is more compact and the 
     coefficients may be readily interpreted physically, by 
     comparison with well-studied solutions of the linearised black 
     hole Einstein equations 
     \cite{ReggeWheeler57,Zerilli70,Chandrasekhar84}.
\end{enumerate}

We next describe the methods used to transform fields from the field value and 
Fourier representations to the spherical harmonic representation.
Transformations for fields of spin~0, 1 and 2 are required  by the code; 
the spin~0 case which we describe here to illustrate the technique is 
slightly less complicated, since we may assume the field $f$ is 
real-valued.

There are $(L+1)^2$ basis functions in the set $\{ Y_{l m}: 0\leq |m| 
\leq l \leq L\}$.  However, to represent these functions as 
trigonometric polynomials on a regular $S^2$ grid we require a grid of 
size $(L+1)\times 2(L+1)$, and thus $2(L+1)^{2}$ real coefficients.  
The spin~0 functions of angular momentum at most $L$ therefore form a 
subspace of real dimension $(L+1)^2$ in the space of Fourier series 
representable on the grid, which has real dimension $2(L+1)^2$.  We 
use a projection onto the spherical harmonic subspace which is 
orthogonal with respect to the natural inner product in the Fourier 
space,
\begin{eqnarray}
\langle f_1,\, f_2\rangle_{2} &=& \frac{1}{4\pi^2}\int_0^{2\pi}\int_0^{2\pi}
f_1(\vartheta,\varphi)f_2(\vartheta,\varphi)d\vartheta d\varphi 
\nonumber
\\
\label{f1f2}
&=& \frac{1}{N^2}\sum_{i=1}^{N}\sum_{j=1}^{N} f_1(\vartheta_i,\varphi_j)
f_2(\vartheta_i,\varphi_j)\,,
\end{eqnarray}
where $\{(\vartheta_i,\varphi_j): i,j=1,\ldots,N\}$ are grid points 
(cf.~\bref{phik}) and $N = 2(L+1)$.

{}To make use of (\ref{f1f2}) we use (\ref{T-torus}) to extend functions 
defined on $S^2$ to functions defined on the torus $\bT^2 = S^1\times S^1$.  
In particular, given any set of values $\{f_{ij} \in \bR: i=1,\ldots,N/2,\ 
j=1,\ldots,N\}$ on the $S^2$ grid, we use (\ref{T-torus}) to construct grid 
values on $\bT^2.$ There is then a unique interpolating trigonometric 
polynomial $f$ such that $f(\vartheta_i, \varphi_j) = f_{ij}$.  We project 
$f$ to the $l\le L$ spherical harmonic subspace as follows.

The $Y_{l m}$ are not orthonormal with respect to (\ref{f1f2}), but 
instead have Fourier inner product
\begin{eqnarray}
     G_{l m\; l^\prime m^\prime} &=& \langle Y_{l m},\, 
Y_{l^\prime m^\prime} \rangle_{\scF} \nonumber \\
&=& \langle \overline{P}_{l m}(\vartheta),\,
          \overline{P}_{l^\prime m}(\vartheta) \rangle_{\scF} \,
   \delta_{m m^\prime}\,.
\label{Gdown}
\end{eqnarray}
Here, the index pair $l m$ (and $l^\prime m^\prime$) is 
a combined index which takes $(L+1)^2$ values, and \bref{Gdown} is the 
matrix for the induced Fourier metric on the spin~0 subspace. 
The summation 
convention will be employed for raised and lowered repeated indices.

For fixed $m$, the inner product (\ref{Gdown}) of the $\overline{P}$ functions 
forms a matrix, 
\begin{displaymath}
           A_{(m)ll^\prime} = \langle \overline{P}_{l m}(\vartheta)
,\, \overline{P}_{l^\prime m}(\vartheta) \rangle_{\scF} \,.
\end{displaymath}
These matrices are defined only for $|m| \leq l,l' \leq L$, so are square 
and of size $(L+1 - |m|)\times (L+1 - |m|)$.  Denoting the inverse matrix
by $A_{(m)}^{ll^\prime}$, we have
\begin{equation} \label{Gdown2}
       G_{l m\; l^\prime m^\prime} =   A_{(m)ll^\prime} 
                                       \delta_{m m^\prime}\,,         
\end{equation}
and the components of the inverse metric are given by
\begin{equation}\label{Gup}
   \quad\quad\quad\quad\quad      G^{l m\; l^\prime m^\prime} 
=   A_{(m)}^{ll^\prime} 
                                       \delta^{m m^\prime} 
\ \quad (\mbox{no sum on}\ m).       
\end{equation}
The dual basis vectors for the spin~0 subspace are 
\begin{equation}\label{Yup}
         Y^{l m} = G^{l m\; l^\prime m^\prime} 
                            Y_{l^\prime m^\prime}\,,
\end{equation}
and satisfy
$
       \langle  Y_{l m},\, Y^{l^\prime m^\prime} \rangle_{\scF}
  =  \delta^{l^\prime}_l \delta^{m^\prime}_m\,
$.
The orthogonal projection of $f$ onto the subspace is given by
\begin{displaymath}
   {\rm proj}(f) = \langle f,\, Y^{l m} \rangle_{\scF} Y_{l m} 
         = f^{lm}Y_{lm},                   
\end{displaymath}
where 
\begin{equation}\label{flm}
                   f^{l m} = \langle f,\, Y^{l m} \rangle_{\scF}
\end{equation}
are the spherical harmonic coefficients of the function $f$.

{}To calculate the inner product (\ref{flm}), first note that using 
(\ref{Ylm}), (\ref{Gup}) and (\ref{Yup}), the dual basis vectors can be 
written as
\begin{equation}\label{Yup2}
             Y^{l m} = \overline{P}^{l m}(\vartheta) F^m(\varphi)\, 
\end{equation}
(in analogy with (\ref{Ylm})), where we have set 
\begin{equation}\label{Pup}
             \overline{P}^{l m}(\vartheta) = 
      A_{(m)}^{l l^\prime} \overline{P}_{l^\prime m}(\vartheta)\,, 
      \quad
    F^m(\varphi) = F_{m}(\varphi) .
\end{equation}
By Fourier analysis of $f$ in the $\varphi$ direction we can write
$f = \hat{f}^k(\vartheta)F_k(\varphi)$. In particular, by $\varphi$-FFT of 
$\{f_{ij}\}$ we get the numbers $\hat{f}^k(\vartheta_i)$. 
The spectral coefficients $f^{lm}$ can then
be evaluated using (\ref{f1f2}) and (\ref{Yup2}) as
\begin{eqnarray}
       f^{l m} &=& \langle \hat{f}^k(\vartheta)F_k(\varphi),\, 
      \overline{P}^{l m}(\vartheta)F^m(\varphi) \rangle_{\scF} \nonumber \\
&=& \langle \hat{f}^m(\vartheta),\, \overline{P}^{l m}(\vartheta) 
\rangle_{\scF} \nonumber \\
&=&  \frac{1}{N} \sum_{i=1}^{N} \hat{f}^m(\vartheta_i)
                       \overline{P}^{l m}(\vartheta_i)\,.
\label{fup}
\end{eqnarray}
The converse process of reconstructing the function values $f_{ij} =
f(\vartheta_i,\varphi_j)$ from the spherical harmonic coefficients
$f^{l m}$ follows from
\begin{eqnarray*}
                  f &=& f^{l m}\, Y_{l m}(\vartheta,\varphi) \\
                    &=& f^{l m} \overline{P}_{l m}(\vartheta)F_m(\varphi)\,.
\end{eqnarray*}
First we construct the quantities
\begin{equation}\label{fhatm}
 \quad\quad\quad\quad\quad 
  \hat{f}^m(\vartheta_i) = \sum_{l=|m|}^L f^{l m} 
                           \overline{P}_{l m}(\vartheta_i)\,,
\ \quad (\mbox{no sum on}\ m),
\end{equation}
and then we use inverse FFTs in the $\varphi$ direction to reconstruct 
$f_{ij}$ via
\begin{displaymath}
   f_{ij} =\sum_{m=-L}^{L} \hat{f}^m(\vartheta_i) F_m(\varphi_j)\,.
\end{displaymath}

Both constructions, of $f^{lm}$ from $f_{ij}$ and conversely, are 
$O(L^{3})$ operations, due to the matrix multiplications in 
\bref{fup}, \bref{fhatm}.  Routines for transforming between grid 
values and spherical harmonic coefficients have been implemented for 
maximum angular momentum $L = 7,15\ {\rm and}\ 31$.  The grid values 
$\overline{P}^{l m}(\vartheta_i)$ which appear in the sum (\ref{fup}) 
were pre-computed in multiple precision using REDUCE \cite{Hearn}. 
The functions 
$\overline{P}^{l m}(\vartheta)$ defined by (\ref{Pup}), were 
constructed symbolically using exact inversion of the matrices 
$A_{(m)\,ll^\prime}$.  This symbolic approach was feasible because the 
metric $G_{l m\; l^\prime m^\prime}$ factorized as the tensor product 
(\ref{Gdown}), thus allowing exact inversion of $G$ using matrices of 
size at most $(L+1)\times(L+1)$ rather than $(L+1)^2\times(L+1)^2$.

The analysis of spin~1 and spin~2 grid functions into spherical 
harmonic coefficients is similar, but complicated by the fact that the 
induced metric on the subspace factorizes as a tensor product only in 
a complex (mixed parity) basis.  Separating the even and odd parity 
coefficients therefore requires some extra book keeping.

Techniques for handling spherical harmonic spectral representations 
have been described by many authors 
\cite{Merilees73b,Orszag74,Swarztrauber79,Swarztrauber89,Jakob97}.  
Our method differs from the Muchenhauer and Daly projection (see 
\cite{Swarztrauber79}) in the choice of inner product (\ref{f1f2}) 
used to define the orthogonal subspace.  More general spectral 
transform methods (eg.  \cite{Swarztrauber89}) use other choices of 
weightings and node points to define the projection, and do not have 
such a simple underlying inner product.  All these methods are also 
$O(L^{3})$.  Jakob \cite{Jakob97} gives an $O(L^{2}\log L)$ spectral 
projection, which however bypasses the construction of the spherical 
harmonic coefficients.  Since we need the spectral coefficients, and 
because we work with a relatively small value of $L$, the Jakob 
projection would not provide any improvement.

The torus method described here and in \cite{Norton96} has the 
advantage that it applies also to higher rank tensors, in particular 
vectors and 2-tensors.  Representations in terms of spin-weighted 
fields are more efficient for vectors (spin $s=1$) than 3-vector 
representations \cite{Swarztrauber81,Swarztrauber84}, and the operator 
$\edth$ gives a transparent derivation of all invariant derivative 
combinations \cite{Swarztrauber81}.

\subsection{Convolution splines}

At various stages it is necessary to interpolate and differentiate
grid-based fields.  For example, the radial integration of the
hypersurface equations by the 8th order Runge-Kutta method requires values 
of the
source field $\beta$ at $10$ intermediate points; the dynamic
regridding of the radial grid requires interpolation to determine the
field values of $\beta$ at the new grid points; and derivatives such as
$\partial\gamma/\partial r$ and $\partial Q^{+}/\partial z$ must be
computed from field values on grids.  A convolution spline algorithm
described in \cite{Norton92} provides a convenient technique.

The method has the effect of fitting a spline curve to sample data, and
is implemented by a convolution of the form \cite{Norton92}
\begin{equation}
    \bar{f}(x) = \sum_{k\in\bZ}f(k)\phi_{n}(x-k),
\label{Cspline}
\end{equation}
where the $f(k)$ are the raw data (samples) and $\phi_{n}(x)$ is a
$C^{n-2}$ sampling kernel. 

The sampling kernel $\phi_{n}$ is constructed as a certain sum of central 
B-splines $M_{n}$ of order $n,$ 
\begin{equation}
        \phi_{n}(x) = \sum_{i=1}^{n-1} a_{i}^{(n)}M_{n}(x-\tfrac{n}{2}+i),
\label{phin}
\end{equation}
where the coefficients $a_{i}^{(n)}$, $i=1,\ldots,n-1$ are chosen so 
that the convolution \bref{Cspline} acts as the identity on 
polynomials $f(x)$ of degree $n-1$ ($n$ even) or $n-2$ ($n$ odd).  
Recall that the central B-spline $M_{n}(x)$ is a $C^{n-2}$ piecewise 
polynomial of degree $n-1$ normalised by $\sum_{k \in \bZ}M_n(x-k) = 
1$, with support on $|x|\le n/2$ \cite{Schoenberg73}.  The support of 
the kernel $\phi_{n}(x)$ is therefore $|x|\le n-1.$
 
\begin{figure}[ht]
        \centering
        \resizebox{8cm}{!}{\includegraphics{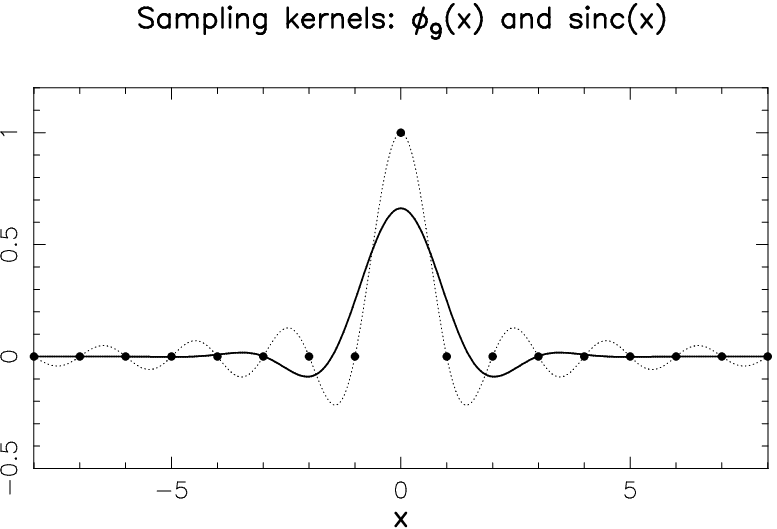}} 
        \caption{Comparison of the $C^{7}$ spline kernel $\phi_{9}$ (solid curve) 
        and the sinc method \cite{Boyd89,Stenger93} kernel 
        ${\rm sinc}(x)= \sin(\pi x)/\pi x$ (dotted curve).  
        The kernel for a sampling method is the 
        response to the delta-like discrete data (solid dots).  The 
        advantages of the convolution spline method are that the kernels 
        have finite support ($[-8,8]$ in this case) and the data is 
        automatically filtered.  The convolution \bref{Cspline} using kernel
        $\phi_{9}$ exactly reproduces polynomials of degree $7$ or less.  }
        \label{fig:phi}
\end{figure}

Algorithms for computing the $a_{i}^{(n)}$ and tabulations for $n\le11$ are 
given in \cite{Norton92}. Coefficients for the kernel $\phi_{9}$ 
used in the code are given in Table \ref{a9i}, and $\phi_{9}$ is 
plotted in Figure \ref{fig:phi}.

\begin{table}[hb]
   \centering
   \caption{Convolution coefficients $a_{i}^{(9)}$.}
   \begin{tabular}{|c|c|c|c|c|}
       \hline
                $i$: & $1,8$ & $2,7$ & $3,6$ & $4,5$  \\
                \hline 
                $a_{i}^{(9)}$:\rule[-2mm]{0mm}{6mm} 
                        & $-\frac{67}{2520}$ & $\frac{1111}{5040}$ 
                        & $-\frac{421}{560}$ & $\frac{1333}{1260}$ \\
                \hline
   \end{tabular}
   \protect\label{a9i}
\end{table}

The expressions \bref{Cspline}, \bref{phin} may be rearranged into a form 
which is more efficient for numerical calculations,
\begin{equation}
        \bar{f}(x) = \sum_{k\in\bZ} \tilde{f}_{k}M_{n}(x-\tfrac{n}{2}-k),
\label{tfk}
\end{equation}
where the modified sample values $\tilde{f}_{k}$ are given by
\[
   \tilde{f}_{k} = \sum_{i=1}^{n-1} f(k+i)a_{i}^{(n)}.
\]
The advantage of \bref{tfk} is that the $\tilde{f}_{k}$ can be 
computed once and then reused to evaluate $\bar{f}(x)$ at many different 
points $x$, using the explicitly known values of the B-spline $M_{n}(x).$ 
The same $\tilde{f}_{k}$ values may also be used to compute 
derivatives of the spline function $\bar{f}$,
\begin{equation}
        \bar{f}^{(j)}(x) = \sum_{k\in\bZ} 
        \tilde{f}_{k}M_{n}^{(j)}(x-k-\tfrac{n}{2}),
\label{dtfk}
\end{equation}
where again the derivatives $M_{n}^{(j)}(x)$ are known functions. These 
techniques are routinely used to supply intermediate values 
and derivatives of fields in the radial and time directions.

Non-uniform distributions of sample points are handled by a mapping between 
the independent variable and the sample number variable ($x$ in the above).
Numerical derivatives are then calculated using the chain rule. 
For example, the radial
grid described in Section \ref{r_grid} is non-uniform, specified by some
known relation of the general form $r = r(n)$ (where $n$ is now being used to 
denote the sample number variable, with radial grid points being at 
$n=0,\ldots,n_\infty$). The operator $\frac{\partial}{\partial r}$ is then 
implemented as $\left(\frac{dr}{dn}\right)^{-1}\frac{\partial}{\partial n}$,
with a formula for the first factor being known explicitly.
  
Similarly, one can transform to an independent variable $s = s_0 + hx$ 
which has grid spacing $h$, to examine the behaviour of the 
approximation \bref{Cspline} as $h \rightarrow 0.$ Let $g(s) := 
f((s-s_0)/h) = f(x)$, so $g^{(j)}(s) = h^{-j}f^{(j)}(x)$.  It can be 
shown \cite{Norton92} that using the $\phi_9$ kernel, the Taylor 
series truncation errors for \bref{Cspline} at a grid point $s$ are
\begin{equation}
        |\bar{g}^{(j)}(s) - g^{(j)}(s)| = c_{j} 
                          h^{8}|g^{(8+j)}(s)| + O(h^9),\quad j=0,1,2,
\label{errj}
\end{equation}
where $c_{0}=\frac{2021}{134400}$, $c_{1}=\frac{4547}{302400}$, 
$c_{2}=\frac{4549}{302400}$.  This reflects the fact that convolution with 
$\phi_{9}$ is exact on polynomials of degree $7$. 
 
\begin{figure}[ht]
        \centering        
        \begin{tabular}{ll}
          \resizebox{0.47\textwidth}{!}{\includegraphics{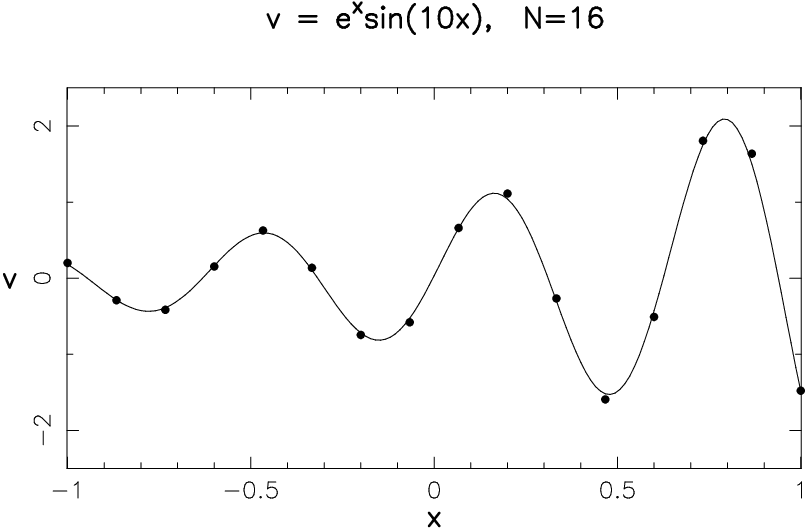}} &
          \resizebox{0.47\textwidth}{!}{\includegraphics{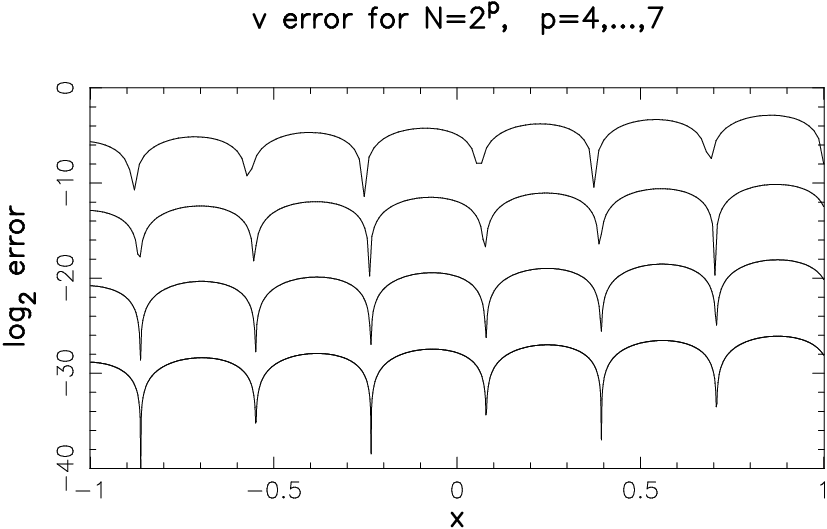}} \\
          (a) & (b)
        \end{tabular}
        \caption{Convergence of the $\phi_9$ spline convolution:
        (a) samples of $v(x) = e^{x}\sin 10x$ at  
         $N=16$ points over $[-1,1]$, and the 
        corresponding convolution spline; (b) logarithmic plots of 
        the absolute error $|v(x) - \bar{v}(x)|$ for grid resolutions
        of $N=2^p$ with $p=4,\ldots,7,$ showing a reduction in the error
        by a factor of $2^8$ on each doubling of the resolution.}    
      \label{v_approx}
\end{figure}

The predicted $h^8$ convergence of the $\phi_9$ spline convolution is clearly
evident
in Figure \ref{v_approx}. Here the function $v(x) = e^{x}\sin 10x$ has been 
approximated at varying grid resolutions corresponding to $N=2^p$ grid points 
over the interval $[-1,1]$. 

Convolution splines do not generally preserve sample values, except 
for samples from polynomials of degree less than or equal to the 
degree of reproduction.  This results in some damping of high 
frequency components of the data, which we expect helps to suppress 
numerical noise and algorithmic instabilities.

Within the context of spectral methods for PDEs, the direct filtering 
of Fourier coefficients of a numerical solution is common practice and 
has been extensively studied (cf.  \cite[\S 8.3]{CanutoEt88} and 
references).  On the other hand, explicit use of a digital filter 
\cite{Hamming77} in conjunction with finite difference methods is 
comparatively rare.  Nevertheless, from an algorithmic point of view, 
this is the effect of using a convolution spline.

The filtering inherent in the method can be examined via the response
function
\begin{equation}
         \Phi_{n}(\theta) = \sum_{k\in\bZ}\phi_{n}(k) \cos k\theta,\quad 
                0 \le\theta\le\pi,
        \label{Phi}
\end{equation}
which is equal to the factor by which the Fourier mode $e^{\imu\theta 
x}$ is amplified by the approximation \bref{Cspline} at a grid point 
$x \in \bZ.$ The value $\theta = \pi$ corresponds to the Nyquist 
frequency for the grid.
 
\begin{figure}[ht]
        \centering
        \resizebox{8cm}{!}{
         \rotatebox{0}{ 
          \includegraphics{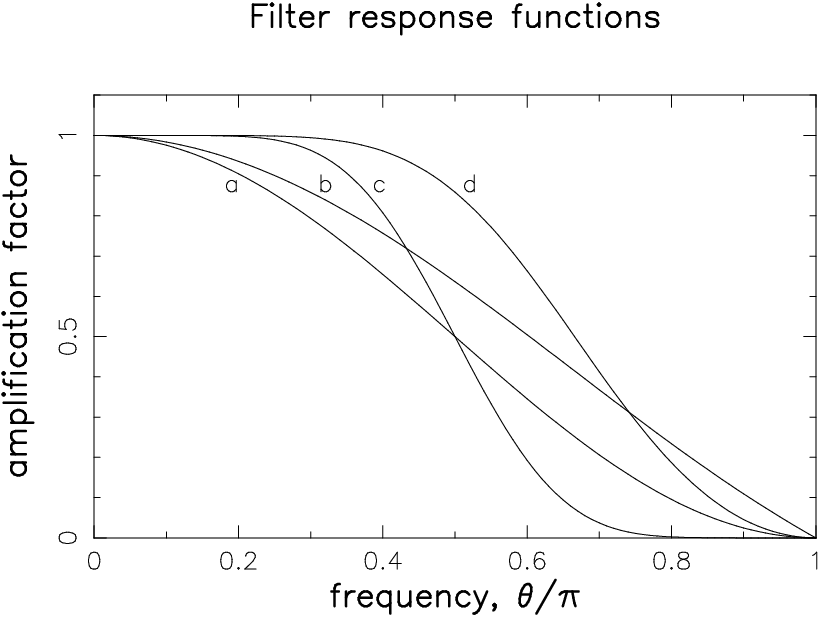}}}
        \caption{Filter response functions:
        (a) raised cosine filter $\sigma(\theta) = \half(1+\cos \theta)$;
        (b) Lanczos filter $\sigma(\theta)=\theta^{-1}\sin\theta$;
        (c) sharpened raised cosine \cite{CanutoEt88};
        (d) $\Phi_{9}(\theta)$.}
        \label{fig:resp}
\end{figure}

Figure \ref{fig:resp} shows the response function $\Phi_{9}(\theta)$, 
compared to the well-known Lanczos and raised cosine (artificial 
viscosity) filters \cite{Hamming77}.  The filtering characteristics of 
convolution splines and their derivatives are described in 
\cite{Norton92}.

\begin{figure}[ht]
        \centering
        \resizebox{\textwidth}{!}{
        $\begin{array}{cc}
          \includegraphics{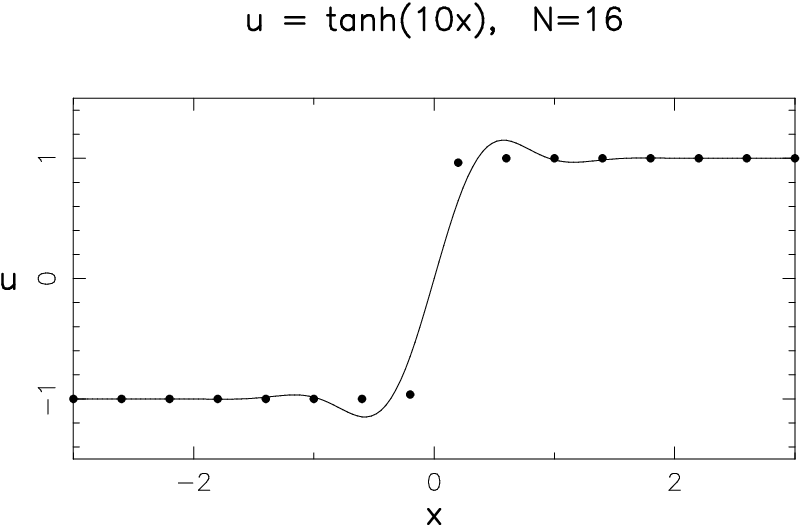} & \includegraphics{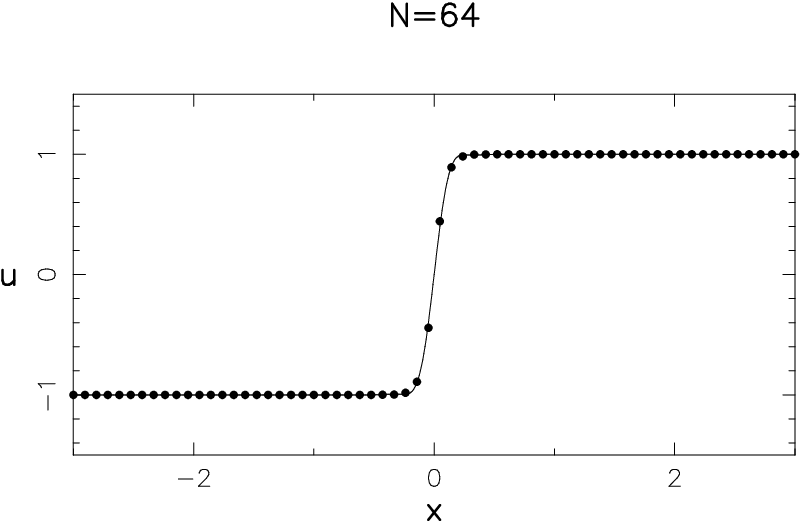} \\
          \includegraphics{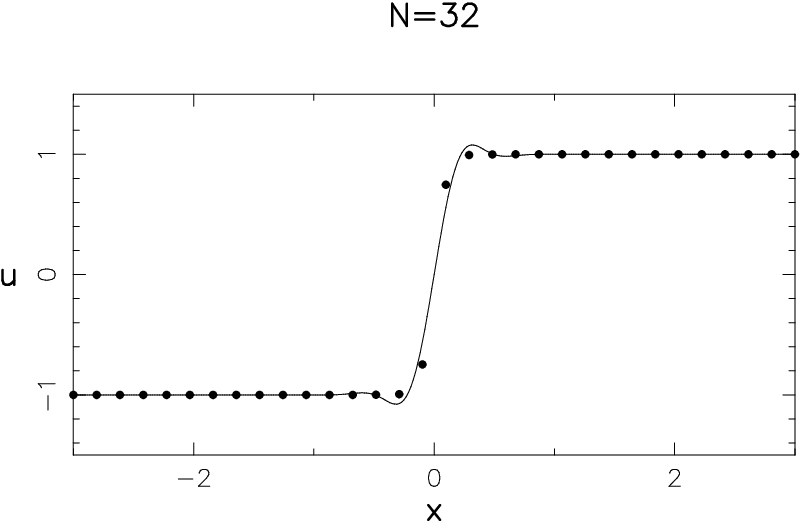} & \includegraphics{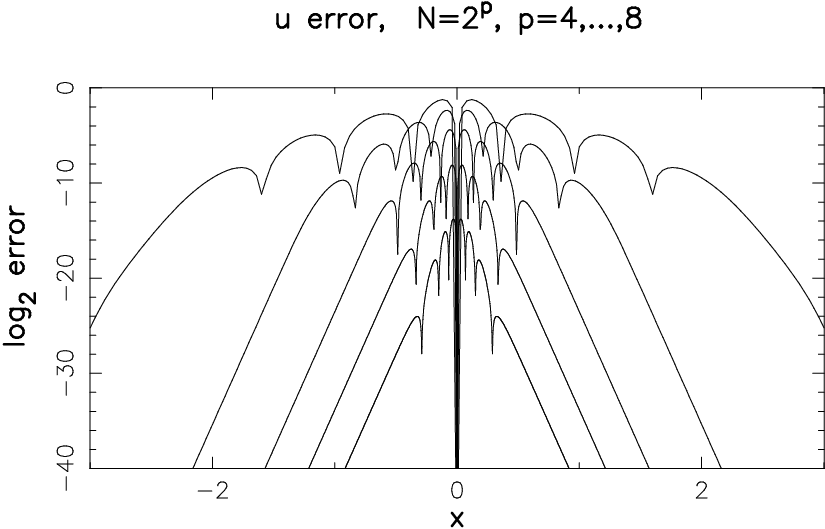} 
        \end{array}$
        }
        \caption{Convolution spline approximation of step-like data. 
        Here the function 
        $u = \tanh(10x)$ is represented by $N$ samples over
        the interval $[-3,3]$. For $N=16$, the transition from 
        $u\approx -1$ to $1$ takes just one grid interval and a Gibbs-like 
        phenomenon
        is evident. The convolution splines are constructed using the 
        $\phi_9(x)$ kernel. The logarithmic plot shows the absolute error, 
        $|u(x) - \bar u(x)|$.       
}
        \label{fig:step}
\end{figure}

The limitations of convolution splines are illustrated in Figure 
\ref{fig:step}, which shows Gibbs-like effects associated with 
approximation of step-like data.  The figures also give an indication 
of the number of grid points needed to resolve a sharp transition in 
field values.

In order that the convolution spline smoothing should introduce only 
negligible errors, the grid resolution should be chosen sufficiently 
fine that typical field variations take place over enough grid points 
that the expected frequency $\theta$ of the field data lies well 
within the part of the Nyquist frequency interval where 
$\Phi_{n}(\theta) \approx 1$.  Of course this requires some prior 
knowledge of the length scale of the field, and cannot be applied 
where shocks (or arbitrarily rapid variations) occur in the data.  In 
such cases the spherical harmonic representation would become equally 
unsuitable.

The choice of high order convolution splines ($h^8$ rather than say 
$h^4$) was motivated by the need to reduce storage requirements.  Low 
order spline convolutions have a markedly reduced usable proportion of 
the Nyquist interval \cite{Norton92}, so to avoid over-smoothing and 
to achieve comparable accuracy with a low order method would require 
significantly higher grid point densities.  Both storage costs and the 
cost of the radial integration increase linearly with the number of 
radial grid points.  The $\phi_9$ kernel was chosen also because its 
accuracy matches that of the RK8 method used for the radial 
integration.

{}To use convolution splines near endpoints of a data set, one can extend 
the data set, using a suitable mapping 
between the independent variable and the sample number variable 
\cite{Norton98}.  The mapping is chosen so that when expressed 
in terms of the sample number variable $n$, $0\le n\le n_{\infty}$, 
the fields admit expansions in powers of $n^{2}$ near $n=0$ and 
$(n_{\infty}-n)^{2}$ near $n=n_{\infty}$.  The sampling kernel 
convolution can then be applied to the even extension of the fields 
through $n=0$ or $n=n_{\infty}$.  This technique is particularly 
important in extracting radiation data near null infinity (scri, 
$\mathcal{I}^{+}$), where the radial grid is chosen so that 
$n_{\infty}-n = O(r^{-1/2})$, as described in Section \ref{r_grid}.

\section{Solution algorithm}  
\label{sec:4}

The hypersurface equations (\ref{eq:Gll}-\ref{eq:Gmm}) suggest the
following process for evolving the 
metric in the exterior region with interior boundary 
the cylinder $r=r_{0}$:
\begin{enumerate}\label{algorithm}
\item  
    Choose boundary data $(H,\Qm,J,K)$ on the cylinder $r=r_{0}$, 
    consistent with the boundary equations 
    (\ref{eq:Gnn},\ref{eq:Gnm});
\item  
    Assume $\beta$ is given on a null hypersurface $\cN_{z}$;
\item  
    Solve the $\cN_{z}$ hypersurface equations 
    $r\partial_{r}U=F(\beta,U)$, by integrating along the radial 
    curves $(z,\vartheta,\varphi)=const.$ with initial conditions at 
    $r=r_{0}$ determined in step 1;
\item  
    reconstruct the metric functions $u,v, \gamma$ from $H,J,K$ and 
    $\beta$ using the converse construction 
    (\ref{u:def}--\ref{v:def});
\item  
    reconstruct $\partial\beta/\partial z$ from $Q$ and the now known 
    values of $\beta,\gamma$ on $\cN_{z}$, using \bref{dbdz:eq};
\item  
    use $\partial\beta/\partial z$ from step 5 to evolve $\beta$ to 
    the ``next'' null hypersurface $\cN_{z+\Delta z}$ and repeat from 
    step 3.
\end{enumerate}
In the following we will show how this heuristic algorithm is
implemented numerically, using the techniques and data representations
of the previous section.

\subsection{Geometry and inner boundary conditions}

\begin{figure}[ht]
        \centerline{ 
          \resizebox{7cm}{!}{\includegraphics{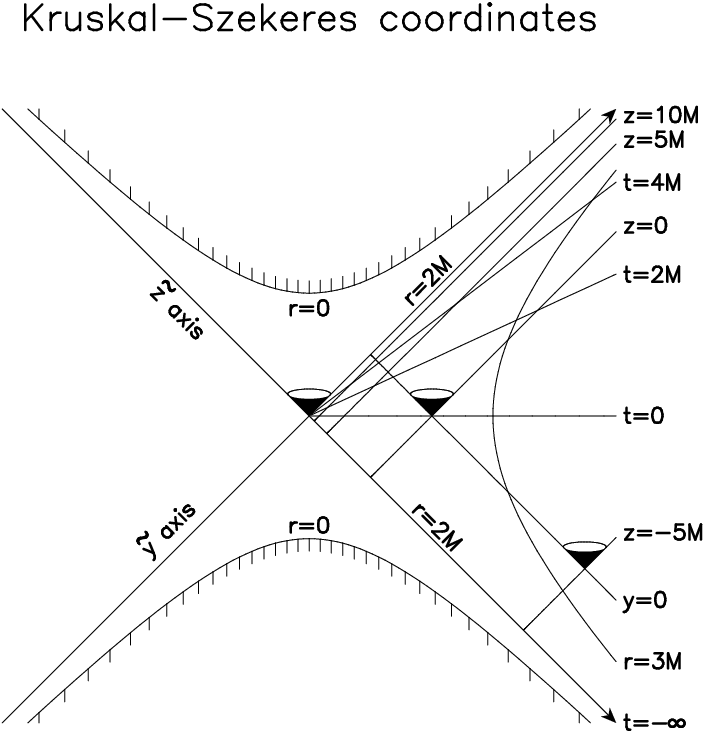}}
        }
        \caption{Schwarzschild spacetime in Kruskal-Szekeres coordinates. 
        Radial light rays are straight lines at $45^{\circ}$.}
        \protect\label{KS:fig}
\end{figure}

\begin{figure}[ht]
        \centering        
          \begin{tabular}{ll}
            \resizebox{!}{7cm}{
                         \includegraphics{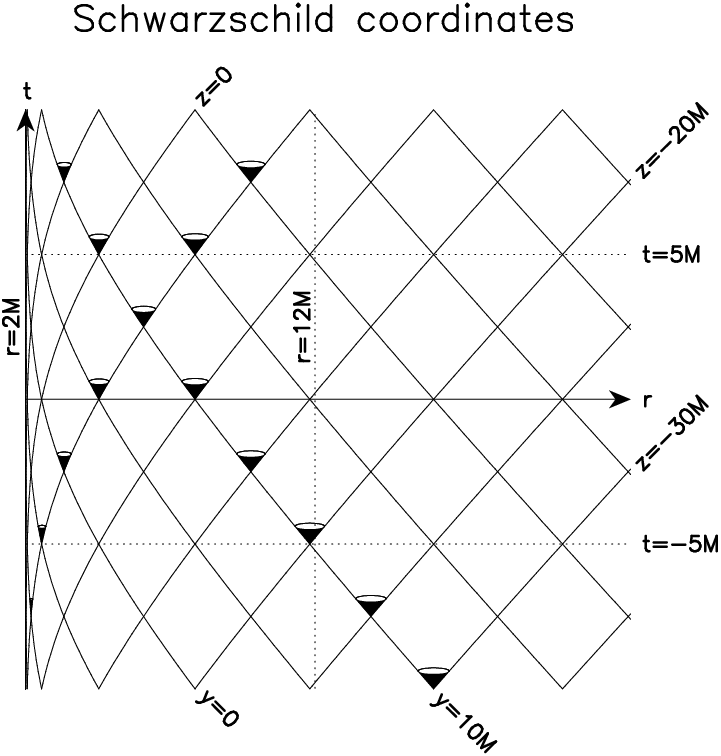}} &
            \resizebox{!}{7cm}{\includegraphics{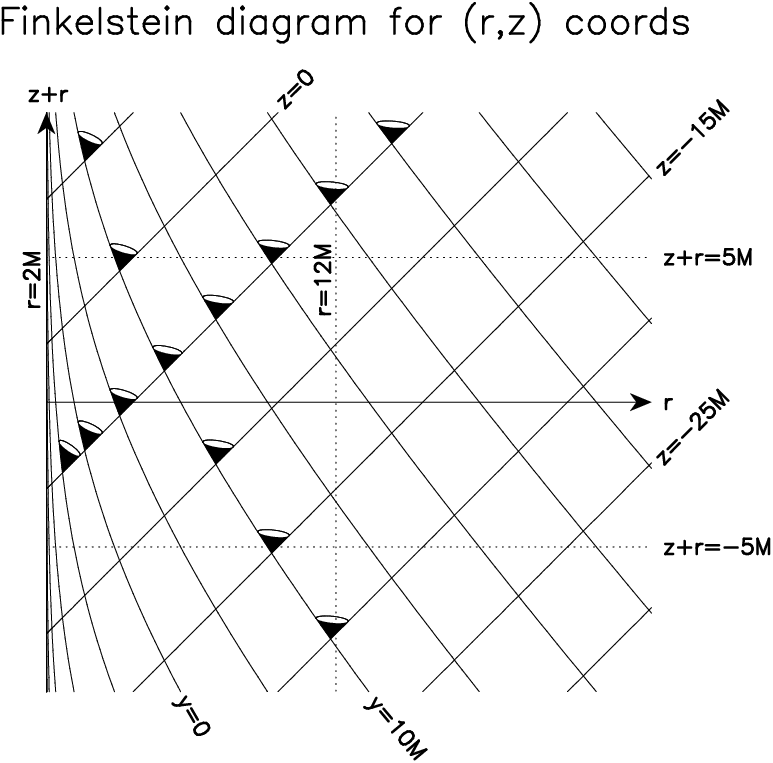}} \\
            (a) & (b)
          \end{tabular}
        \caption{Exterior region $r>2\mschw$ of Schwarzschild spacetime, with 
        past and future radial null geodesics. (a) In Schwarzschild 
        coordinates $(r,t)$; (b) in retarded Eddington-Finkelstein
        coordinates $(r,z).$}
        \label{rtschw:fig}
\end{figure}

The code models gravitational waves propagating on a black hole 
spacetime, with metric approximating that of the Schwarzschild 
solution in the Kruskal-Szekeres coordinates \cite{HawkingEllis73}.  
Introducing the double-null coordinates
\[
        z=t-r^{*},\qquad y=t+r^{*},\qquad 
        r^{*}=r+2\mschw\log\left(\frac{r}{2\mschw} - 1\right), 
\]
the Schwarzschild
metric becomes
\[   ds^{2}_{Schw} = -(1-2\mschw/r)\,dy\,dz + r^{2}d\Omega^{2}\,,  \]
where $d\Omega^{2}= d\vartheta^{2}+\sin^{2}\vartheta\,d\varphi^{2}$.
The coordinate singularities at the past and future horizons
$t=\pm\infty$ are removed by defining $\tilde{y}=e^{y/4\mschw}$,
$\tilde{z}=e^{-z/4\mschw}$, giving the metric
\[
   ds^{2}_{Schw} =
     \frac{32\mschw^{3}}{r}e^{-r/2\mschw}\,d\tilde{y}d\tilde{z} +
     r^{2}\,d\Omega^{2}\,,
\]
where $r=r(\tilde{y},\tilde{z})$ is defined implicitly by (see Figure 
\ref{KS:fig})
\begin{equation}
      e^{r/2\mschw}\left(\frac{r}{2\mschw} - 1\right) = 
      \tilde{y}\tilde{z} .
 \label{tilde_yz}
\end{equation}
Because radial light rays are straight 
lines at $45^{\circ}$ and $r=0$ is singular, the surfaces $\tilde{z}=0$ 
and $\tilde{y}=0$ (ie.  $r=2\mschw$) form the past and future event 
horizons, and these are smooth hypersurfaces with bounded curvature.  
The approximate Minkowski structure of Schwarzschild spacetime is 
better illustrated by the radial null geodesics in $(r,t)$ 
coordinates, see Figure \ref{rtschw:fig}.  Note however that the 
$(r,t)$ coordinates are singular along the event horizons $r=2\mschw$.


Initial conditions for $\beta$ are imposed on $\{z=0,r\ge 2\mschw\}$ 
(with $\mschw=1$ usually), by specifying the spherical harmonic 
coefficient functions $\beta_{lm}(r)$.  Since $\beta(z=0)$ is unconstrained, 
these coefficient functions may be freely chosen, subject only to the 
size condition \bref{bsize}.

For simplicity the inner boundary conditions are set at 
$r_{0}=2\mschw=2$ to agree with the Schwarzschild past horizon: 
$H_{0}=2$, $Q_{0}=J_{0}=K_{0}=0$.  Since we choose $\beta(0,r)=0$ for 
$2\le r \le 5$, by causality the solution should agree with the 
Schwarzschild metric in a neighbourhood of $r=2$ for all time $z\ge0$, 
producing a ``white hole'' past horizon in the spacetime.  This choice 
of inner boundary condition has the considerable advantage that the 
boundary equations (\ref{eq:Gnn}),(\ref{eq:Gnm}) are automatically 
satisfied, and it is not necessary for this class of simulations to 
separately ensure that the boundary data are numerically compatible 
with the boundary equations.

The resulting spacetimes have the geometry of an isolated black hole with 
future event horizon at $z=\infty$, $r>2\mschw$ and Schwarzschild-like white 
hole boundary along $r=2\mschw, 0\le z <\infty$.  Adding shear $\beta$ at the 
initial hypersurface $z=0$ results in spacetimes modelling the interaction 
of gravitational radiation with a single black hole.

Although the fixed past horizon boundary conditions used in the 
present code allow many interesting issues to be addressed, it would 
be desirable to implement more general inner boundary conditions.  
Such conditions specify $H,J,K,\Qp$ at an inner surface ($r=1$, for 
example), subject to the dynamical ($\partial/\partial z$) constraints 
on the evolution of $J/u$ and $\Qp$ determined by the boundary 
equations \bref{eq:Gnn},\bref{eq:Gnm} \cite{Bartnik97a}.  The free 
data on the inner boundary consist of $u_{0},K_{0}$, where $u_{0}$ 
represents a certain coordinate gauge freedom, whilst $K_{0}$ 
describes the gravitational radiation injected into the system through 
the inner boundary.  Various exact solutions with such boundary 
conditions are described in \cite{Bartnik97b} (Robinson-Trautman 
\cite{RobinsonTrautman62}, boosted Schwarzschild, twisted Minkowski 
space), and would provide useful accuracy checks on the numerical 
methods.

However, implementing general inner boundary conditions raises 
numerical and geometric difficulties --- the boundary data must be 
``consistent'' with the RK4 solution evolution in order to maintain 
optimal accuracy (see \cite{AbarbanelEt96} for an analysis of similar 
but simpler situations), and constraining the radiation data $K_{0}$ 
such that the spacetime is still Schwarzschild near the past horizon 
is a difficult geometric problem.  An arbitrary choice of $K_{0}$ 
(even $K_{0}=0$ if $u_{0}\ne 1$) will inject some additional energy 
into the spacetime.  

\subsection{Dynamic radial grid}
\label{r_grid}

There are two geometric features which the code should model 
accurately: future null infinity (``scri'' or $\scri^{+}$, where 
$r\to\infty$, $z$ finite), and the future horizon $r\sim 2\mschw$, 
$z\to\infty$.  

The field near null infinity $\scri^{+}\cap \cN_{z}=(r=\infty, z)$ 
determines the outgoing gravitational waves as seen by a distant 
observer and is consequently very important for applications to 
gravitational wave astronomy.  Experience with 3+1 codes shows that it 
is not possible (as yet) to provide boundary conditions on an outer 
timelike boundary at a finite radius which do not either inject 
radiation or reflect radiation back into the grid.  This deficiency 
has the effect of severely limiting the overall time duration of most 
$3+1$ simulations.  We avoid all such reflection problems by using a 
radial grid coordinate $n$, which compactifies $r=\infty$ and leads 
to accurate modelling of gravitational radiation.

The $(z,r)$ coordinates become singular near the future horizon 
$z\to\infty$ in the Schwarzschild spacetimes (see Figures \ref{KS:fig}, 
\ref{rtschw:fig}).  In our case this picture is not exact, since the 
spacetime geometry is only approximately Schwarzschild, and thus the 
future horizon (for example) will not be located exactly at 
$z=\infty$.  However, the NQS parameterisation must still become 
singular eventually, as the outgoing null hypersurfaces $\cN_{z}$ 
approach the future event horizon.

The effect of the nearly singular coordinates is that at late times, 
the in-falling gravitational components near the event horizon will be 
compressed into a region of small $r$-variation, and this compression 
will accelerate in time $z$, whilst retaining field structures from 
early times.  Consequently no $r$-grid which is constant in time is 
able to accurately represent the in-falling radiation at late times.  
We have observed that numerical problems with a fixed radial grid 
arise as early as $z=10$.

{}To overcome these problems, a dynamic and variable radial grid is 
used, based on double null coordinates $(z,\tilde{y})$.  The time 
steps in the evolution direction are regular, with $\Delta 
z=0.1,0.05,0.025$ being typical.  The grid in the radial
direction is chosen to satisfy the criteria that it compactify null 
infinity and concentrate grid points in the region of greatest 
variation in the seed field $\beta$.  Because the field features 
propagate along the inward and outward null characteristics, which 
correspond respectively to the curves $\tilde{y}=const.$ 
(approximately!)  and $z=const.$ (exactly), the numerical grid is 
taken to be rectangular in the $(z,\tilde{y})$ coordinates, with 
radial grid point positions being determined by an initial 
distribution of grid points on the surface $z=0$.

Introducing the radial grid coordinate $n$ with range $0\le n\le 
n_{\infty}$ (with typical values of $n_{\infty}$ being 128, 256, 512 and 
grid points at integer $n$), we specify an  
initial grid point distribution $r(z=0,n)=f(n)$, 
where $f$ is some monotone increasing function such that 
$f(n_{\infty}) = \infty$. 
The radial grid points on the initial ($z=0$, $\tilde{z}=1$) surface have 
$\tilde{y}$ ordinates given by \bref{tilde_yz},
\begin{equation}
        \tilde{y} = (f(n)/2\mschw -1)\exp({f(n)/2\mschw}) = 
        \phi(f(n)/2\mschw),
        \label{tyn}
\end{equation}
where $\phi(x) := (x-1)e^{x}$ is monotone and invertible for $x\ge0$.  
Since $\tilde{y},z,r$ are related by 
$\tilde{y}=e^{z/4\mschw}\phi(r/2\mschw)$ and the grid points are 
required to inflow along the curves of constant $\tilde{y}$, we can 
determine the dynamic radial grid point distribution $r=r(z,n)$ in 
terms of the initial grid distribution function $f(n)$ by
\begin{equation}
   r(z,n) = 2\mschw\, 
   \phi^{-1}\!\left(\exp({-z/4\mschw})\phi(f(n)/2\mschw)\right),
\label{eq:r(z,n)}
\end{equation}
where the inverse function $\phi^{-1}:[-1,\infty)\to[0,\infty)$ is 
evaluated numerically.  With this definition, the surfaces $n=const.$ 
correspond to in-falling null hypersurfaces in the reference 
Schwarzschild metric.  In order to express the hypersurface equations 
in terms of $n$ rather than $r$, we need to compute $\partial 
r/\partial n$ --- this expression follows immediately from 
\bref{eq:r(z,n)}:
\begin{equation}
        \frac{\partial r}{\partial n} = e^{-z/4\mschw} \frac{f(n)}{r(z,n)}
            \exp\left({\frac{f(n)-r(z,n)}{2\mschw}}\right) \frac{df}{dn}.
\label{eq:drdn}
\end{equation}

\begin{figure}[ht]
        \centerline{ 
          \resizebox{10cm}{!}{\includegraphics{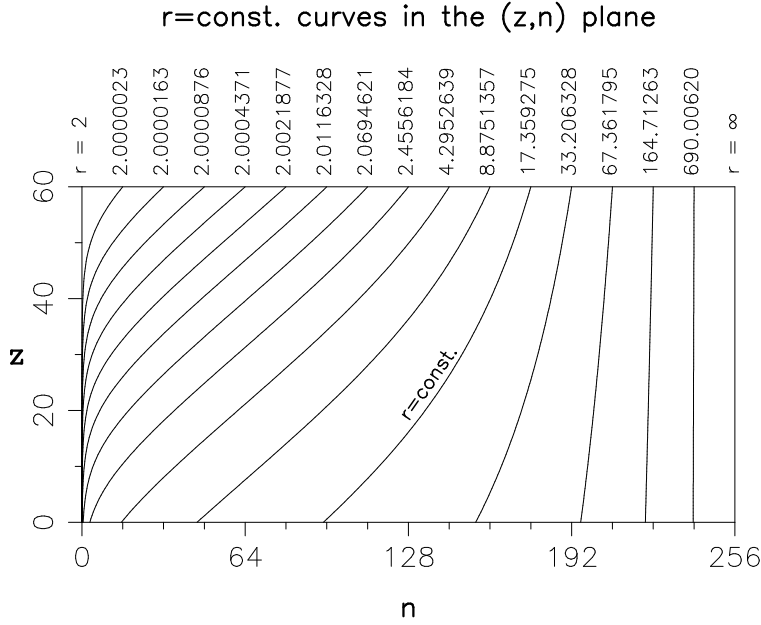}}
        }
        \caption{The map between radial grid point number $n$ and radius $r$ 
        is dynamic, chosen so that grid points approximately follow  
        inward null geodesics. At late null-time $z,$ grid points cluster 
        near the black hole horizon at $r=2.$ 
        }
        \protect\label{fig:rzn}
\end{figure}

\begin{figure}[ht]
        \centerline{ 
          \resizebox{8cm}{!}{\includegraphics{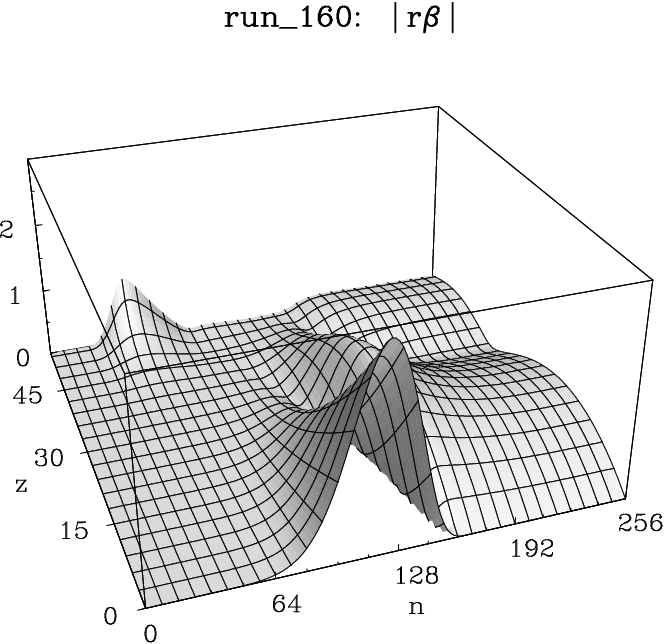}}
        }
        \caption{Evolution of $r\beta$ for $0\le z \le 55$, in radial in-falling 
        coordinates.  Observe that the in-falling grid tracks the dynamical 
        evolution.  This simulation has timestep $\Delta z=0.05$, 
        $n=0$ is the past horizon $r=2\mschw$ and
        $n=256$ represents future null infinity $\scri^{+}$.}
\protect\label{fig:beta}
\end{figure}

It remains to choose the initial grid distribution $f(n)$.  The 
condition $n_{\infty}-n = O(r^{-1/2})$ is achieved by setting $f(n) = 
f_{1}(\nu)/(1-\nu)^{2}$, where $\nu=n/n_{\infty}$ and $f_{1}:[0,1]\to\bR$ is 
any suitable smooth monotone bounded function.  In the code, $f_{1}$ is a 
quadratic polynomial, with coefficients chosen to concentrate grid points 
across the support of the chosen initial data $\beta(z=0)$.  Figure 
\ref{fig:rzn} shows sample curves $r(z,n)=const.$, illustrating the 
in-falling nature of the $(z,n)$ grid coordinates.
This heuristic prescription for distributing the grid points works 
well in practice --- Figure \ref{fig:beta} shows the shear over the 
$(z,n)$ plane for \texttt{run\_160}, and clearly demonstrates the 
in-falling structure of this solution.  The simulation eventually 
terminates at $z = 55$ because of some geometric effect associated 
with breakdown of the NQS gauge condition near the future event 
horizon.

\subsection{Hypersurface equations}

The hypersurface equations are solved by treating them as a large system of 
ordinary differential equations, with the radial grid coordinate $n$ playing 
the role of independent variable, and the dependent variables being the 
values taken by the fields $(H,J,Q,K)$ at the $N^2/2$ points of the 
$(\vartheta,\varphi)$-grid.

The form (\ref{eq:lh}--\ref{eq:K}) of the hypersurface equations, for the 
variables $\log H$, $r\Qp$, $j$ and $K$, proves to be better behaved near 
$r=\infty$, since each of these variables has a finite (usually non-zero) 
limit.  Integration of these radial ODEs is possible up to and including 
the final point $n=n_{\infty}$, with results whose numerical effectiveness 
may be seen by inspecting the field values in a neighbourhood of null 
infinity \cite{gular}.  Tests described in the following section, 
in particular the consistency of the constraint equations 
and the accuracy of the Trautman-Bondi 
mass decay formula (Figure \ref{fig:massloss}) also confirm that asymptotic 
behaviour has been reliably calculated.

Note that unlike methods based on Bondi-Sachs or Newman-Unti 
coordinates \cite{Bondi62,NewmanUnti63}, integration along the $r$-coordinate 
lines does not correspond to integrating along the radial null 
geodesics (the characteristics of the Einstein equations), since in 
general the NQS shear $\beta$ is non-zero, and the null direction is 
$\ell=\partial/\partial r -r^{-1}\beta$.

The radial integration with respect to the coordinate $n$ is performed 
using an 8th order Runge-Kutta scheme \cite{DormandPrince81}, with RK 
step size $\Delta n = 1$.  The RK8 method requires 13 derivative 
evaluations per RK step, of which 10 are at intermediate points not on 
the radial grid.  Values of the field $\beta$ and its angular 
derivatives at these intermediate points are provided by convolution 
splines generated using the kernel $\phi_{9}$ with samples at integer 
$n$.

\subsection{Reconstructing the metric}  
\label{sec:elliptic}

Step 4 of the solution algorithm requires us to reconstruct the metric 
functions $(u,v,\gamma)$ from the solution $(H,J,K,Q)$ of the 
hypersurface system (\ref{eq:Gll}--\ref{eq:Gmm}) with seed $\beta$ and 
boundary data $(H,J,K,Q)|_{r=r_{0}}$.  Note that the connection 
variables (\ref{def:H}--\ref{def:Qpm}) are determined by the values of 
the metric functions $(\beta,\gamma,u,v)$ and $\partial \beta/\partial 
z$ on the hypersurface $\cN_{z}$.

The reconstruction is carried out as described in section 
\ref{sec2.3}.  This process requires solving the system 
(\ref{Lbeta:def}) on each $S^{2}$ of the radial grid.  If $\beta$ is 
not too large, then (\ref{Lbeta:def}) is an elliptic system of partial 
differential equations on the sphere $S^{2}$, mapping surjectively to 
the space of spin~2 fields.  We solve (\ref{Lbeta:def}) by first 
substituting
\begin{equation}
        \gamma = \edth^{-1}\Gamma,
\label{def:Gamma}
\end{equation}
where $\edth^{-1}$ is defined spectrally by
\[ \edth^{-1} Y^{2}_{lm} 
= - \left[\half(l+2)(l-1)\right]^{-1/2} Y^{1}_{lm},\quad l\ge 2, \] 
so (\ref{Lbeta:def}) becomes
\begin{equation}\label{Gammeq}
      \cK_{\beta}\Gamma :=     
      \Gamma + \frac{\edth \beta}{2-\div\beta}\div(\edth^{-1}\Gamma)
\;=\; -K + J \frac{\edth \beta}{2-\div\beta} \,. 
\end{equation}
Note that the choice \bref{def:Gamma} gauges the $l=1$ spherical 
harmonic components of $\gamma$ to zero --- a similar but more 
expensive construction may be used if nonzero $\gamma_{l=1}$ 
components are desired.  The advantage of (\ref{Gammeq}) over 
(\ref{Lbeta:def}) is that the operator $\cK_{\beta}$ in (\ref{Gammeq}) 
is close to the identity for small $B:= \edth\beta/(2-\div\beta)$.  
The corresponding discretized problem is therefore well 
suited to iterative matrix methods.

We use the conjugate gradient (CG) method \cite{Johnson87}, which is 
an iterative method applicable to matrix problems of the form $Ax = b$ 
with $A$ symmetric positive definite.  Accordingly we actually solve 
an associated self-adjoint equation obtained by applying to 
(\ref{Gammeq}) the operator adjoint to that in (\ref{Gammeq}) with 
respect to the $L^2$ norm on $S^2$.

In order to compute the adjoint operator $\cK_{\beta}^{T}$, notice first that 
$\cL_{\beta}$ and $\cK_{\beta}$ are {\em real}-linear but not 
complex-linear, so the adjoint must be computed with respect to the real 
form of the inner product (\ref{s2ip}).
Expanding $\Gamma=\sum_{l\ge2}\Gamma^{lm}Y^{2}_{lm}$, 
$\phi=\sum_{l\ge1}\phi^{lm}Y^{1}_{lm}$ we have the spectral representations
\begin{eqnarray}
\label{edthinv}
        \edth^{-1}\Gamma & = & 
        -\sum_{l\ge2,m}\left[\half(l-1)(l+2)\right]^{-1/2}\Gamma^{lm}Y^{1}_{lm},
\\
\label{divpspec}
        \div\phi & = & \sum_{l\ge1,m}\left[\half 
        l(l+1)\right]^{1/2}(\phi^{lm}+\bar{\phi}^{lm})Y_{lm},
\end{eqnarray}
and thus the real adjoints $(\edth^{-1})^{T}$, $\div^{T}$ are 
\begin{eqnarray}
\label{edthinvT}
        \edth^{-1T}\phi & = & -\sum_{l\ge2,m}
           \left[\half(l+2)(l-1)\right]^{-1/2}\phi^{lm}Y^{2}_{lm},
\\
\label{divT}
        \div^{T}(f+\imu g) & = & 2 \sum_{l\ge1,m}\left[\half l(l+1)\right]^{1/2}
            f^{lm}Y^{1}_{lm},
\end{eqnarray}
where $f,g$ are real-valued functions.  Consequently we may represent the 
adjoint $\cK_{\beta}^{T}$ spectrally by
\begin{equation}
        (\cK_{\beta}^{T}\phi)^{lm} = \phi^{lm} - \sqrt{\frac{l(l+1)}{(l-1)(l+2)}}
           (\bar{B}\phi + B\bar{\phi})^{lm},
\label{KT}
\end{equation}
where $B=\edth\beta/(2-\div\beta)$ and $l\ge2$.  

The equation (\ref{Lbeta:def}) transformed into (\ref{Gammeq}) gives 
 an invertible equation
\begin{equation}
        \cA_{\beta}\Gamma := \cK_{\beta}^{T}\cK_{\beta}\Gamma = 
\cK_{\beta}^{T}(JB-K).
\label{AGamma}
\end{equation}
The right hand side of \bref{AGamma} may be computed explicitly using 
the spectral representation and \bref{KT}.  The operator $\cA_{\beta}$ 
is symmetric and positive definite and close to the identity when 
considered in the spectral representation.  Thus the conjugate 
gradient algorithm may be applied to \bref{AGamma} and we see that the 
parameterisation of \bref{Lbeta:def} in terms of $\Gamma$ 
\bref{def:Gamma} amounts to a preconditioner.  Note CG requires not 
that the \emph{matrix} of $\cA_{\beta}$ be given explicitly, but only 
that $\cA_{\beta}\Phi$ can be evaluated for any spin~2 field $\Phi$.  
We carry out this evaluation by a sequential process which computes 
the actions of $\cK_{\beta}$ and $\cK_{\beta}^{T}$ using the spectral 
representations of $\edth^{-1},\edth^{-1T}$ and $\div,\div^{T}$ (which 
are simple diagonal operators) combined with transformations to the 
point representation to evaluate multiplication terms like 
$\bar{B}\Phi$ followed by projections back to the spectral 
representation.

This scheme requires several transformations between representations 
of fields by their spin-weighted spherical harmonic coefficients and 
by their values on the $S^2$ grid.  For example, the operator 
$\edth^{-1}$ is a trivial multiplicative operator on spectral 
coefficients, whereas the products in the source terms are best 
calculated in the grid representation.  Although evaluating the action 
of $\cA_{\beta}$ is thus numerically expensive, the expense is more 
than compensated for by the rapid convergence of the CG algorithm with 
this spectral preconditioning.

The spectral representation has the further advantages that the 
solution is represented fewer unknowns ($2(L+1)^{2}-8$ compared to 
$4(L+1)^{2}$ for the grid value representation), and gauge conditions 
which specify the $l=1$ components of $\gamma$ (eg.~$\gamma_{l=1}=0$) 
can be directly implemented.  It is possible to adapt the algorithm to 
allow for other NQS gauges (eg.~$\beta_{l=1}=0$), but this is 
numerically more expensive since (\ref{Lbeta:def}) must be solved 4 
times at each sphere rather than once.  Although such gauges have some 
geometric advantages \cite{Bartnik97b}, their numerical implementation 
has not yet been considered.

Using CG to solve for the spin~2 spherical harmonic coefficients of 
$\Gamma$ turns out to be quite efficient, typically requiring fewer 
than 10 iterations for an $S^2$ grid of size $N/2\times N = 16\times 
32$.  On this size grid we resolve all components of $\Gamma$ up to 
angular momentum $L = N/2 - 1= 15,$ so in this case we are solving for 
$2((L + 1)^2 - 4) = 504$ spectral coefficients.
The scheme's effectiveness is due in part to having a good initial guess for 
$\Gamma$ to use as the starting point of the CG iterations, namely the 
solution found for $\Gamma$ on the 2-sphere at the previous radial position.

The CG iterations finish when the error, measured by the sum of 
squares of spherical harmonic coefficients of the difference of the 
two sides of \bref{Gammeq}, is $10^{-2}$ times the size of the 
aliasing error in the source term.  This aliasing error is the 
difference between the raw field values of the source term (which is 
necessarily calculated in the field value representation because it 
involves products and quotients) and its field values after projection 
into the subspace spanned by spin~2 spherical harmonics.  It provides 
an estimate of the error in the source term, and hence (because the 
operator $\cK_{\beta}$ is close to the identity) it is reasonable to 
accept a solution of comparable accuracy.

{}To ensure termination of the CG algorithm, other stopping criteria 
are also checked, but the relative error test is the usual termination 
cause and is found to work well in practice.  For example, it can 
result in a 10-fold improvement over letting the CG iterations run 
until the solution is determined to machine precision.

\subsection{Evolution}  
\label{sec:evolution}

Given $\beta$ on a null hypersurface $\cN_{z}$, we construct the time 
derivative $\partial\beta/\partial z$ by solving the hypersurface equations
with seed $\beta$, determining $\gamma,v$ as outlined in the previous 
section, and then using formula (\ref{dbdz:eq}) to evaluate 
$\partial \beta/\partial z$.
Let us write the result of this process as 
\begin{equation} 
        \frac{\partial \beta}{\partial z} = \cB(\beta,U_{0})
\label{BbetaU0}
\end{equation}
where the operator $\cB$ is determined  
by the value of $\beta$ on the hypersurface $\cN$ and the initial conditions
$U_{0}=(H_{0},Q_{0},J_{0},K_{0})$ at $r=r_{0}$
for the hypersurface equations.

The evolution formula (\ref{BbetaU0}) provides the basis of the 
spacetime evolution algorithm, which simply incorporates 
(\ref{BbetaU0}) into a standard 4th order Runge-Kutta algorithm.  This 
approach is just the method of lines, treating the evolution equations 
as a very large system of ordinary differential equations for the 
(point) representation of the entire field $\beta(z) = 
\beta_{|\cN_{z}}$.

The method of lines, applied blindly in this manner, is generally 
prone to instabilities.  Tests suggest the relative stability of the 
NQS code derives from the smoothing effects (a) of the convolution 
spline, and (b) of the spectral projection.  The filtering implicit in 
the convolution spline is applied to $\beta$ during the radial 
integration of the hypersurface equations, at each of the 4 stages of 
the RK4 algorithm.  It is not possible to turn off this radial 
filtering because the convolution splines for $\beta$ are an essential 
part of the algorithm for evaluating the right hand side of 
\bref{BbetaU0}.
 
Smoothing of $\beta$ in the angular directions is done explicitly, by 
projecting $\beta$ onto the spin~1 subspace with maximum angular 
momentum $L$ or $2L/3$ (the Orszag $2/3$ rule, to eliminate quadratic 
aliasing).  This angular filtering is done after each of the 4 stages 
of the RK4 algorithm.  Removal of the angular filtering results in 
very rapid disintegration of the evolution, which then typically lasts 
only a few RK4 steps.

For simplicity, the 4 RK4 stages evolve $\beta$ in the $z$ direction 
in the $(z,r)$ coordinates, along $r=const$.  At the end of each full 
RK4 time step the key field $\beta$ is interpolated onto the new 
radial grid \bref{eq:r(z,n)}, using a convolution spline based on 
values of $\beta$ on the old grid.

The RK4 time integration of $\beta$ evolves field values on the 
$(\vartheta,\varphi)$-grid.  Equivalently, we could have evolved its 
spherical harmonic coefficients, of which there are only half as many.  
However, the amount of computation saved by doing so is insignificant 
in comparison to that required {}to evaluate $\partial\beta/\partial 
z$, so this choice is made for convenience.

Likewise, the RK8 radial integration of the system of hypersurface 
equations uses the field value representation.  In this case, however, 
it is found that projecting the fields onto their appropriate 
spherical harmonic subspaces during the integration is not required 
for either stability or accuracy.  There is a definite computational 
advantage in staying within the field value representation, since 
several relatively expensive $O(L^3)$ projections are avoided.

The first radial derivative of $\gamma$ is needed to evaluate 
$\partial\beta/\partial z$ \bref{dbdz:eq}.  Grid values of 
$\partial\gamma/\partial r$ are calculated numerically as derivatives 
of convolution splines (in the radial direction) for the 
$(\vartheta,\varphi)$-grid values of $\gamma$, making use of formula 
\bref{eq:drdn} and the chain rule for derivatives.  The radial 
derivative term $\cD_r\log u$ appears in the hypersurface equations 
\bref{eq:rQp} and \bref{eq:j}.  Using equation \bref{eq:lh} and 
definition \bref{u:def}, this term can be written as an expression 
involving only the 1st radial and angular derivatives of $\beta$.

The program is normally run until the solution ceases to be well 
behaved.  Blowup is detected by monitoring $2-\div\beta$, which must 
remain everywhere positive.  For the initial data that we have used, 
the blowup has always occurred in the $l=2$ modes of $\beta$, at low 
values of $n$ corresponding to $r-2\mschw\approx 0$ (see Figures 
\ref{fig:rzn},\ref{fig:beta}).  Although the precise cause of the 
blowup is not yet understood, it is not a numerical instability, since 
it is unaffected by changes in the radial or timestep resolutions, nor 
does it appear to be primarily  geometric, 
since most curvature scalars remain bounded.  This 
suggests the blowup is a coordinate effect, probably arising from 
proximity to the future event horizon.

For smooth initial data of intermediate strength (\texttt{run\_160}), 
the evolution extends {}to $z\sim 55$.  The final time varies with the 
strength of the initial data --- see Table \ref{table:datasize}.  The 
evolution of an intermediate strength solution is shown in Figure 
\ref{fig:beta}, which plots the mean square or $L^{2}(S^{2})$ size of 
$\beta$ at each radial sphere, for time $0 \le z \le 55$.  Termination 
is caused by the blowup feature at low radius, which grows steadily 
from time $z=40$ onwards.

\section{Accuracy tests}  
\label{sec:accuracy}

The complexity of the NQS Einstein equations and the variety of 
algorithms employed in the code, make it problematic to prove 
rigorously that the numerical simulation accurately models the physics 
and geometry of the spacetime.  Instead we rely on a range of tests to 
justify the reliability of the code, probing the numerical accuracy 
of the solutions through their convergence and geometric consistency.

We consider here tests based on the \emph{numerical convergence} of 
the solutions as algorithmic parameters are varied; and on the 
\emph{algebraic consistency} of the numerical solutions.  The 
consistency tests measure the constraint identities and the 
Trautman-Bondi mass decay formula \cite{Trautman58b,Bondi62}.  Work in 
progress considers other tests, including comparisons with linearised 
theory, and with better known solutions such as Robinson-Trautman, 
Schwarzschild and Minkowski spacetimes in twisted NQS coordinates 
\cite{Bartnik97b}.

The resolution of the simulations is determined by three parameters:
the spherical harmonic spectral limit $L$ (or effective limit $l_{\rm 
max}$); the number of radial zones $n_{\infty}$; and the time step 
$\Delta z$.  We shall examine in turn how the accuracy of a solution 
depends on each of these parameters.

It is clear that numerical convergence can be estimated from the 
convergence properties of the key field $\beta$.  However, convergence 
of $\beta$ guarantees only that the (limit) solution satisfies \emph{some} 
system of equations, which may not coincide with the desired vacuum 
Einstein equations.  (For example, the Einstein equations may have 
been incorrectly implemented).  Thus, to assert that the correct 
equations have been solved, it is essential to provide independent 
tests of the correctness of the code.

The most natural independent test is to compare the numerical solution 
with an explicitly known solution.  Unfortunately the Schwarzschild 
metric \bref{ds2:schw} is trivial in the NQS gauge and does not 
provide a useful comparison test, whilst the twisted shear-free 
metrics \cite{Bartnik97b} require boundary conditions which are more 
general than those available in the present version of the code.  

Instead we consider here another class of independent tests based on 
constraint relations.  Such relations are typical of geometric 
equations arising in geometry and physics, which admit gauge and 
coordinate freedoms.  Thus, we check the geometric consistency of the 
solution by evaluating $r^2G_{nn}$ and $r^2G_{nm}$, using the 
constraint relations \bref{eq:Gnn} and \bref{eq:Gnm}.  Neither of 
these relations is used in generating the numerical solutions, and in 
theory these components should evaluate to zero.  In practice, since 
each is a sum of terms having magnitude approximately $|\beta|\sim1$, 
the extent to which $r^{2}G_{nn}$, $r^{2}G_{nm}$ evaluate to zero 
serves both to confirm the consistency of the numerical solution with 
the vacuum Einstein equations, and also to assess the accuracy of the 
solution.

The Trautman-Bondi mass decay formula provides another such test of 
geometric consistency, and of the accuracy of the solution near 
$r=\infty$.  This theoretical result leads to a relation between the 
asymptotic ($r=\infty$) values and $z$-derivatives of the fields $H$, 
$J$, and $K$, and may be readily tested for our numerical solutions.

In the following we discuss numerical solutions using three reference 
initial $\beta(z=0)$ fields, which differ only by the scale factors 
given in Table \ref{table:datasize}.  In each case the initial $\beta$ 
consists of pure $l=2,m=2$ spherical harmonics with equal strength odd 
and even parts, and radial profile given by a bump supported on $5\le 
r\le 40$.  We use the terms \emph{weak}, \emph{intermediate} and 
\emph{strong} to describe solutions generated using the three sizes of 
initial data.

Another convenient measure of the strength of the gravitational field 
is the initial relative mass difference $m_{B}(0)/\mschw - 1$, between 
the initial Bondi mass of the numerical spacetime (cf.~\bref{def:mB}) 
and the background Schwarzschild mass ($\mschw$).  Table 
\ref{table:datasize} gives the initial relative mass differences for 
the three reference initial $\beta$ fields.

\begin{table}[ht]
 \centering
 \caption{Size of initial data sets }
 \vskip 3mm
 \begin{tabular}{|c|c|c|c|}
   \hline
     Field strength: & weak & intermediate & strong  \\
   \hline
     $\beta(0)$ scale factor:   & 1 & 4.48 & 10  \\
   \hline
     $m_{B}(0)/\mschw - 1$: & $0.9472\times 10^{-2}$ & $ 0.1915 $
                             & $ 0.9940 $  \\
   \hline
     Last $z$:  &  61 & 55 & 51 \\
   \hline
 \end{tabular}
 \label{table:datasize}
\end{table}

The qualitative conclusions of the error analysis of this section may 
be summarised as follows:

The major determining factor in the overall accuracy is the spectral 
limit $L$.  Truncating spherical harmonic coefficients beyond $L$ has 
the effect of modifying the Einstein equations to a system for which 
the constraint identities are no longer valid, and thereby places a 
lower bound on the numerical accuracy.  For $L=15$ the weak and 
intermediate field solutions can be adequately resolved, but this is 
not sufficient to obtain adequate (beyond $10^{-3}$) accuracy for the 
strong field simulation \texttt{run\_170}.

To suppress unstable quadratic aliasing effects, it is essential to 
use an Orszag 2/3 rule truncation.

Within the bounds governed by the spectral limit $L$, accuracy can be 
improved by increasing the radial resolution $n_{\infty}$.  For the 
weak field solution, $n_{\infty}=1024$ reduces the radial error 
contribution to the level of the spectral truncation error (see 
Figure \ref{Gnn_dz}(b)).

For given resolutions $L$ and $n_{\infty}$, there is a range of values 
$\Delta z$ for which the simulation remains stable.  Outside this 
range, the simulation follows the standard solution for some time, 
then rapidly blows up.  The simulation is largely insensitive to the 
value of $\Delta z$ within the stable range, so $\Delta z$ may be 
chosen as large as possible, consistent with stable evolution.

\subsection{Dependence on spectral limit $L$}
Using our current hardware it is not generally feasible to run the 
code at $L=31$, and $L=7$ is too low to be of interest.  The code is 
normally run at $L=15$ resolution (giving a $16\times 32$ 
$(\vartheta,\varphi)$ grid) with an anti-aliasing cutoff at $l_{\rm 
max}=10$.

Orszag \cite{CanutoEt88, Orszag71} observed that quadratic aliasing 
can be eliminated by periodically removing the upper $1/3$ of the 
spectral bandwidth of a numerical solution.  If fields contain only 
modes for which $l \leq \frac{2}{3} L$, then a quadratic product is 
band limited to $l \leq \frac{4}{3} L$.  With a working bandwidth $L$, 
the modes for which $L \leq l \leq \frac{4}{3} L$ become aliased onto 
the modes $\frac{2}{3}L \leq l \leq L$.  Therefore, truncation at 
$l_{\rm max} = \frac{2}{3}L$ will remove quadratic aliasing 
contamination.

If no cutoff is used (ie.~the full $L=15$ resolution is retained) 
then the high $l$-modes of the intermediate strength simulations blow 
up at $z \approx 8$.  The onset of instability (time until blow up) of 
the $L=15$ simulations with no $l_{\rm max}$ cutoff is largely 
independent of the time step and the radial resolution.  This suggests 
that the effect of the nonlinear aliasing contamination is best 
regarded as changing the system of equations into a system which has 
unstable solutions.

Because the nonlinear interactions in the NQS equations are 
predominantly quadratic, it is not surprising that the $l_{\rm 
max}=10$ cutoff is sufficient for long term stability.  The 
intermediate strength solution lasts until $z=55$, when the code 
terminates for other reasons.

\begin{figure}[h!tb]
   \centering
   (a) \resizebox{0.4\textwidth}{!}{\includegraphics{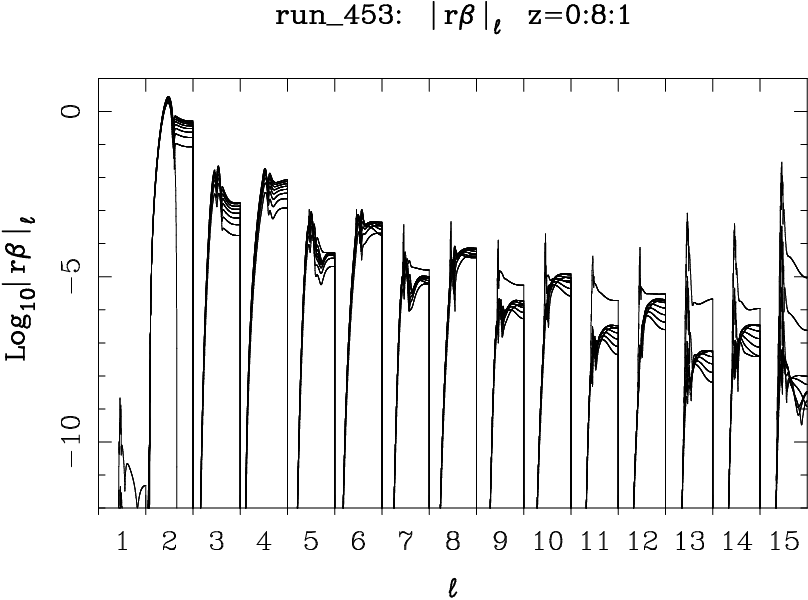}}
   (b) \resizebox{0.4\textwidth}{!}{\includegraphics{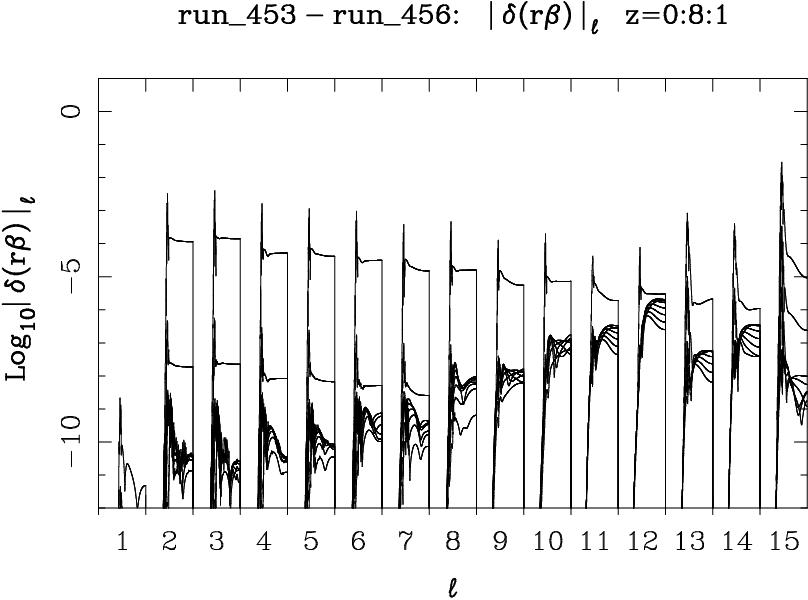}}
   \caption{Orszag's $2/3$ rule is used to remove aliasing instability:
      (a) unstable evolution of the high $l$-modes of an $L=15$ simulation
          (no anti-aliasing cutoff);
      (b) difference between an unstable $L=15$ simulation (no cutoff) 
          and a stable simulation with an $l_{\rm max}=10$ cutoff.   
      Each $l$-bin contains a radial plot (linear in $n$, with $n = 
      0,\ldots,n_\infty$) of the square root of the sum of the 
      squares of the $(l,m)$-components for fixed $l$ with 
      $m=-l,\ldots,l$.  These simulations have $n_\infty = 512$ and 
      $\Delta z = 0.05$.}
\label{aliasing}
\end{figure}

Figure \ref{aliasing}(a) shows blow up of \texttt{run\_453}, an $L=15$ 
simulation of the intermediate field strength solution.  The $l=15$ 
modes show rapid growth beyond $z=6$, indicating the instability 
of the aliasing feedback.  Figure \ref{aliasing}(b) shows the 
difference between \texttt{run\_453} and the stable simulation 
\texttt{run\_456}, which has an $l_{\rm max}=10$ cutoff.  Until the 
onset of the high $l$-mode instability (ie.~for $z \leq 6$) there is 
good agreement between the two simulations, with approximately 
$10^{-10}$ relative difference for $l=2$ modes and $10^{-2}$ relative 
difference for $l=10$ modes.

\begin{figure}[h!tb]
        \centering
        (a) \resizebox{0.4\textwidth}{!}{\includegraphics{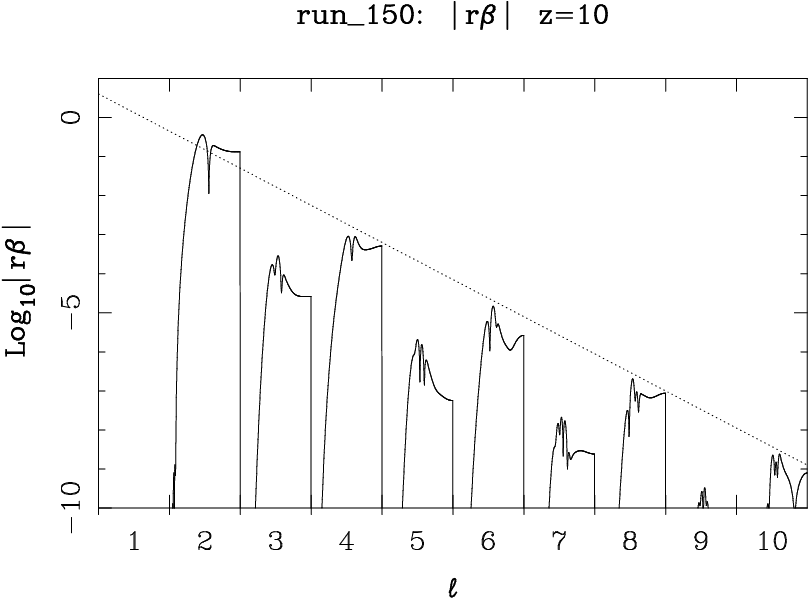}}
        (b) \resizebox{0.4\textwidth}{!}{\includegraphics{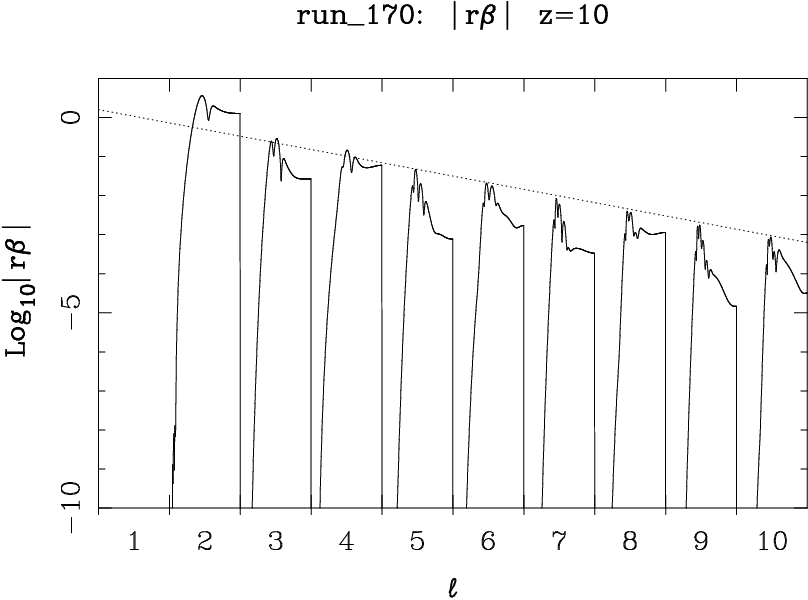}}
        \caption{Spectral resolution and field strength:
        (a) well resolved weak field with fast $l$-mode decay; 
        (b) poorly resolved strong field with slow decay of $l$-modes 
        (see Table \ref{table:datasize} for field strength details).}
        \label{beta_spec}
\end{figure}

The spectral limit $L$ is critical in determining the relation between 
gravitational field strength and simulation accuracy.  This can be 
appreciated by observing the decay rate of the $l$-spectrum of 
$\beta$, as in Figure \ref{beta_spec}.  By extrapolation, the error 
introduced by the anti-aliasing cutoff at $l_{\rm max}=10$ should be 
no more than the $l=10$ coefficient, and a relative error estimate 
follows by comparing the $l=10$ and the $l=2$ coefficients.

Figures \ref{beta_spec}(a) and \ref{beta_spec}(b) show the dramatic 
difference in decay rates of the $l$-modes of $\beta$ for weak and 
strong fields.  Assuming an $l_{\rm max}=10$ cutoff, it is evident 
that the relative error is at most $10^{-8}$ for weak field 
simulations, and about $10^{-3}$ for strong field simulations.

From the observed decay rate of the $l$-modes of $\beta$ for a given 
field strength, it is possible to estimate the resolution $L$ required 
to achieve a prescribed accuracy.  Thus although we cannot directly  
investigate the behaviour of errors with varying spectral limit $L$ 
(due to hardware constraints), we can still investigate spectral 
resolution effects by altering the $\beta$ field strength.

\begin{figure}[h!tb]
        \centering
        (a) \resizebox{0.4\textwidth}{!}{\includegraphics{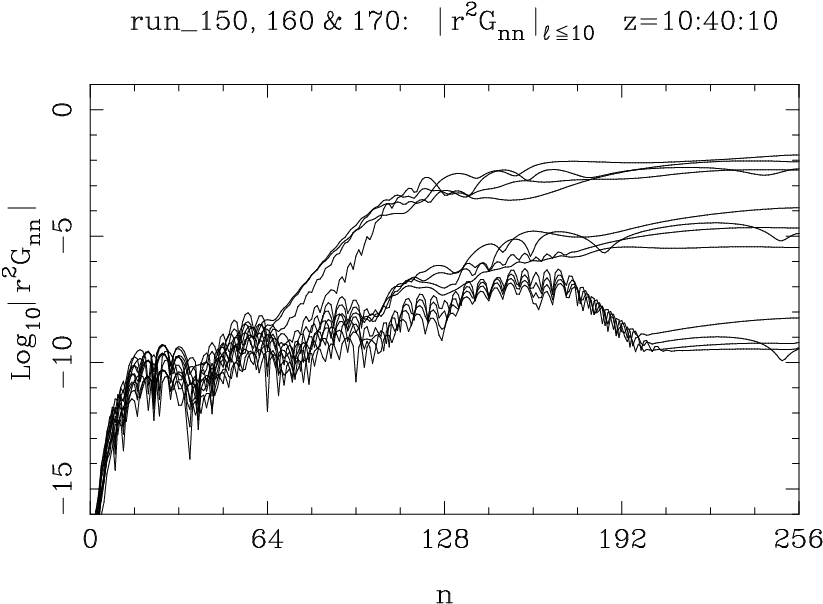}}
        (b) \resizebox{0.4\textwidth}{!}{\includegraphics{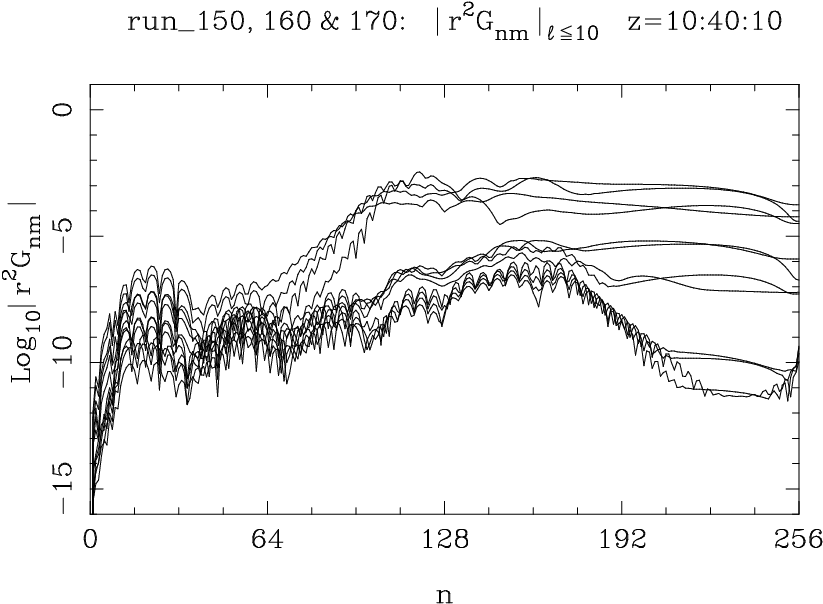}}
        \caption{Effect of spectral resolution on constraint quantities 
        (a) $|r^2G_{nn}|_{S^2}$;
        (b) $|r^2G_{nm}|_{S^2}$,
        at times $z = 10,20,30,40$, for strong (top 4 curves), 
        intermediate (middle 4 curves) and weak (bottom 4 curves) fields.}
        \label{Gxx}
\end{figure}

Figures \ref{Gxx}(a) and \ref{Gxx}(b) show the effect of field 
strength (weak, intermediate, strong) on the constraint quantities 
$r^2G_{nn}$ and $r^2G_{nm}$. The parameters for these simulations are 
$l_{\rm max}=10$, $n_\infty = 256$ and $\Delta z = 0.05$.
The four curves in each band are times $z=10, 20, 30, 40$. There is no 
significant $z$ dependence of either $G_{nn}$ or $G_{nm}$ until within 
about $5\mschw$ of the final blow up time.

The second Bianchi identity implies the conservation law 
$G_{ab}^{;b}=0$, which leads to a radial system of equations for 
$G_{nn}, G_{nm}$ with sources linear in the hypersurface Einstein 
tensor components $G_{\ell\ell}$, $G_{\ell m}$, $G_{\ell n}$, 
$G_{mm}$.  Thus $G_{nn}, G_{nm}$ give a measure of the accumulated 
error in the hypersurface equations in the radial direction.  This 
provides some explanation of the structure of the $G_{nn}, G_{nm}$ 
graphs, particularly for the strong field solution: the numerical 
solution of the hypersurface equations will have greatest error in the 
region where the fields are strongest, in this case the range $64 < n 
< 128$, and this is precisely the region of greatest increase in 
$G_{nn}, G_{nm}$.

\subsection{Dependence on radial grid resolution $n_\infty$}

\begin{figure}[h!tb]  
   \centering
   \resizebox{0.4\textwidth}{!}{\includegraphics{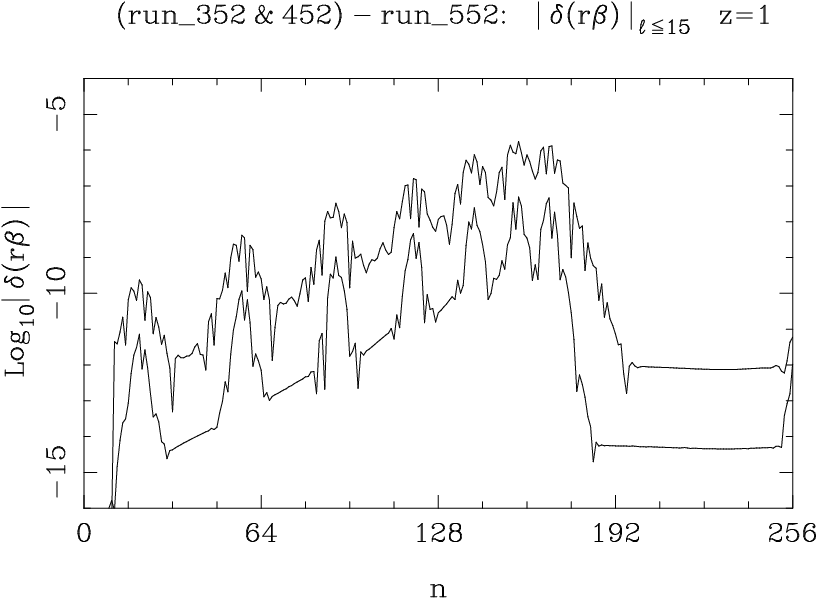}}
   \caption{Convergence of $\beta$ with increasing radial 
        resolution: weak field solutions with $n_{\infty}=256,512$ 
        compared to $n_{\infty}=1024$.  The error decreases by 
        approximately a factor of $2^{8}$ on doubling the radial 
        resolution.}
\label{beta_n}
\end{figure}

The radial regridding and interpolation of $\beta$, the 
radial differentiation of $\gamma$, and the RK integration of the 
hypersurface equations are all formally 8th order accurate. 
Figure \ref{beta_n} shows that this is consistent with the observed 
convergence of $\beta$ on increasing the radial resolution.

\begin{figure}[h!tb]
   \centering
   (a) \resizebox{0.4\textwidth}{!}{\includegraphics{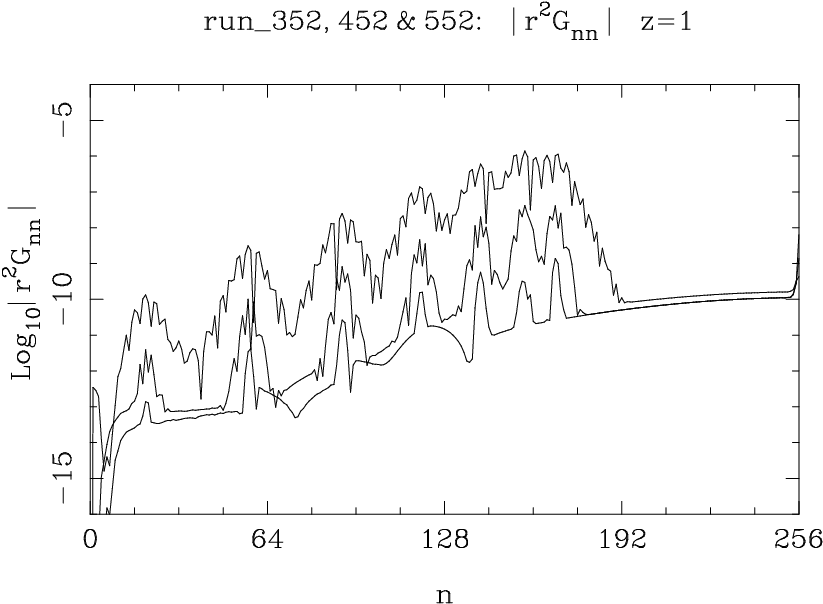}}
   (b) \resizebox{0.4\textwidth}{!}{\includegraphics{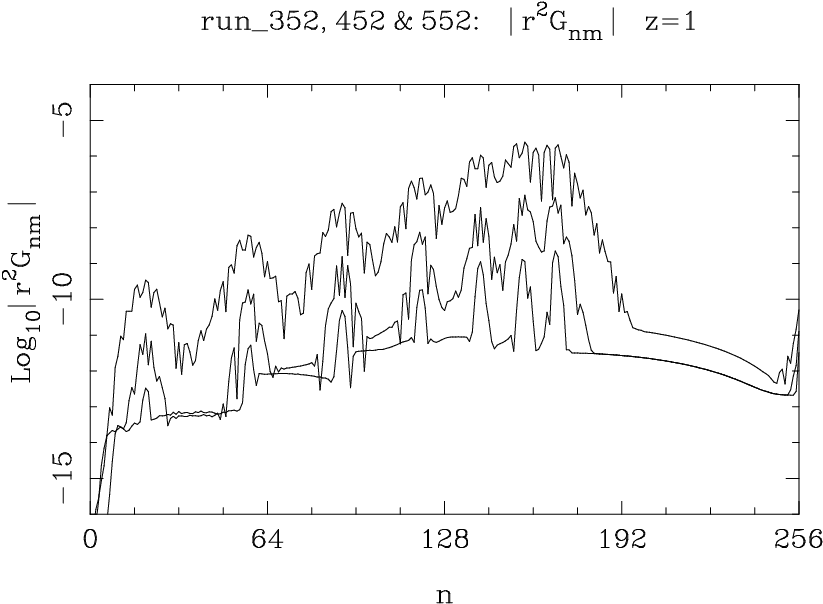}}
   \caption{Effect of radial resolution on weak field constraint 
        quantities (a) $|r^2G_{nn}|_{S^2}$, (b) $|r^2G_{nm}|_{S^2}$.
        In each case the three curves are $n_\infty = 256,512,1024$ 
        (top, middle and bottom curves respectively).}
   \label{Gxx_n}
\end{figure}

The constraint quantities $G_{nn}$ and $G_{nm}$ also exhibit some 
convergence effects.  Figures \ref{Gxx_n}(a) and \ref{Gxx_n}(b) show 
show significant improvement between $n_{\infty}=256$ and $512$, but 
little between $512$ and $1024$.  The form of the $n_{\infty}=1024$ 
curve indicates that errors at the highest radial resolution are 
dominated by errors associated with the spectral truncation.

\subsection{Dependence on time step $\Delta z$}

\begin{figure}[h!tb]
        \centering
        (a) \resizebox{0.4\textwidth}{!}{\includegraphics{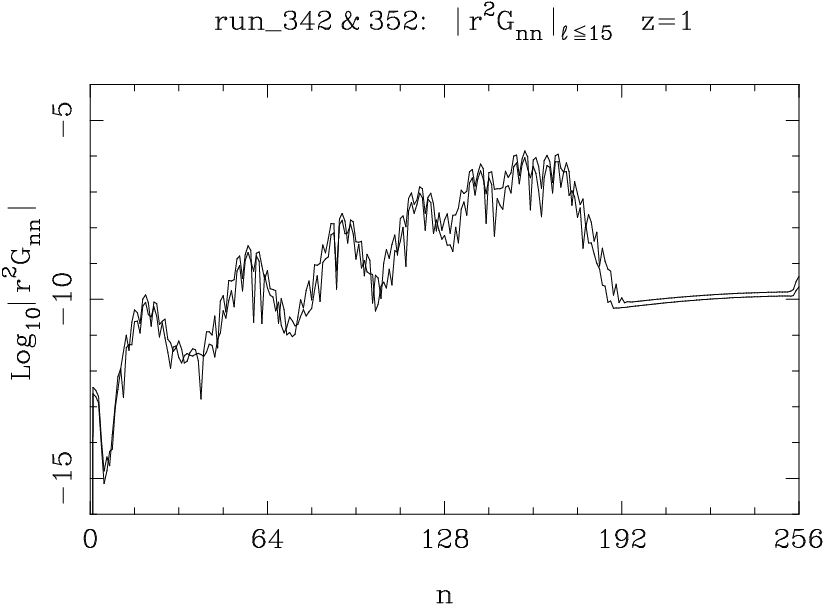}}
        (b) \resizebox{0.4\textwidth}{!}{\includegraphics{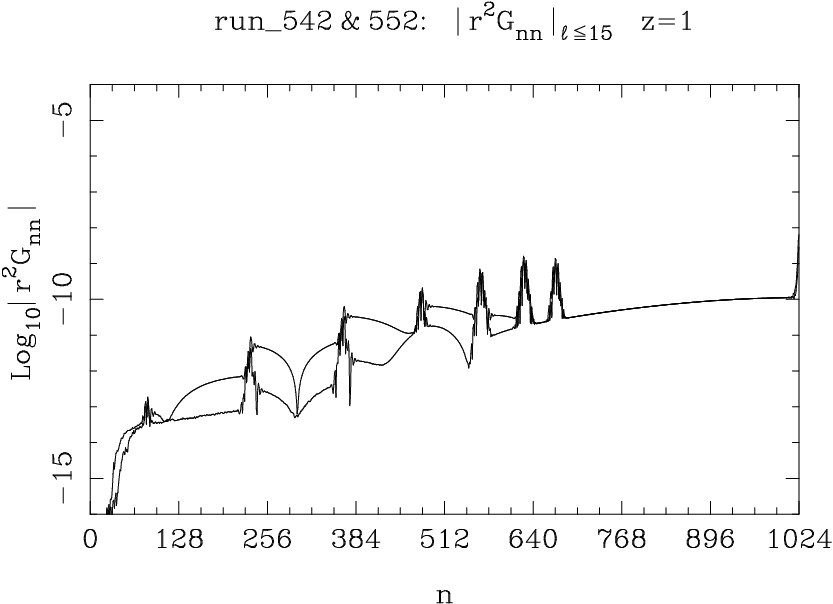}}
        \caption{Effect of time step resolution on the constraint 
            quantity $|r^2G_{nn}|_{S^2}$ for the weak field solution:
        (a) $n_\infty = 256, \Delta z=0.1,0.05$: the error is dominated by 
            the radial discretisation error for $n < 192$ and by the 
            spectral truncation error for $n > 192$. Refining $\Delta z$ 
            produces no appreciable improvement in the solution.
        (b) $n_\infty = 1024, \Delta z=0.1,0.05$: the radial discretisation
            error is small enough that the RK4 integration error can be 
            observed. For $n > 700$ the error is dominated by the spectral 
            truncation, resulting in the same tail as in (a), while for
            $n < 700$ the constraint improves in places, 
            consistent with a factor of 16 decrease in the error.}
        \label{Gnn_dz}
\end{figure}

\begin{figure}[h!tb]
   \centering
   \resizebox{0.4\textwidth}{!}{\includegraphics{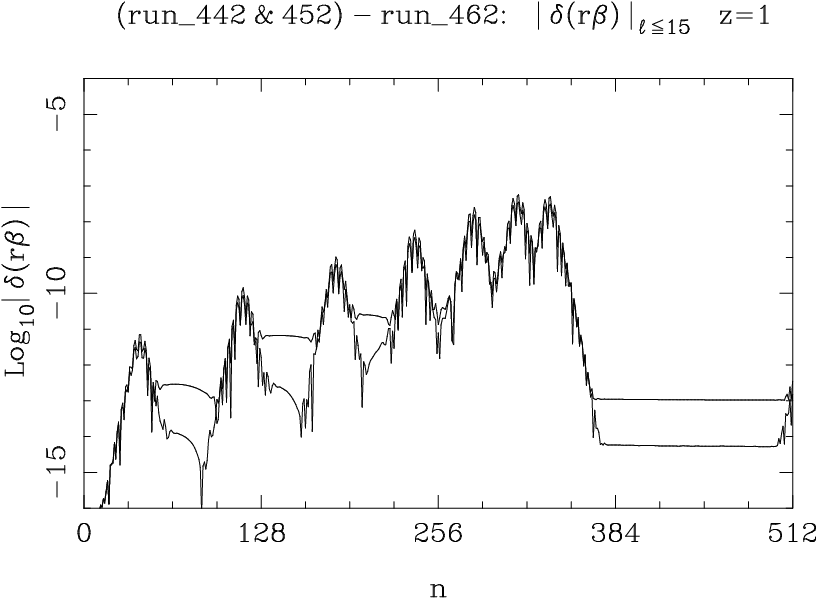}} 
   \caption{Convergence of $\beta$ with decreasing time step: weak 
      field solutions for $\Delta z=0.1,0.05$, compared against 
      $\Delta z=0.025$.  Where the error is not dominated by the 
      radial discretisation error, the curves show a decrease in error 
      which is consistent with 4th order convergence.}
   \label{beta_dz}
\end{figure}

Although the solution algorithm is formally 4th order accurate in the 
time direction, at the typical resolutions at which the code is run, 
the RK4 errors are completely dominated by errors arising from the 
spectral truncation $L$ and/or the radial discretisation $n_\infty$.  
This is illustrated by Figure \ref{Gnn_dz}(a), which shows no 
significant difference in the constraint quantity $G_{nn}$ between 
$\Delta z = 0.1$ and $\Delta z=0.05$, with $n_{\infty}=256$.  However, 
when the solution is better resolved in the radial direction, a small 
effect can be observed, cf.~Figure \ref{Gnn_dz}(b), where 
$n_{\infty}=1024$.  Figure \ref{beta_dz} compares $r\beta$ for runs 
with $\Delta z = 0.1,0.05,0.025$ and $n_{\infty}=512$ and again shows 
only minor improvements from decreasing $\Delta z$.

Consequently, $\Delta z$ is optimally chosen as large as possible, 
subject to resulting in stable evolution.  For $n_\infty = 256$ and 
$L=15$ with an anti-aliasing $l_{\rm max}=10$ cutoff, the evolution 
is stable for $\Delta z= 0.1$ and unstable for $\Delta z= 0.2$, which 
blows up at time $z=25$, after 125 RK4 steps.

\subsection{Energy and asymptotic decay tests}
\label{sec:BTmassloss}

\begin{figure}[h!tb]
  \centering
     (a) \resizebox{0.4\textwidth}{!}{\includegraphics{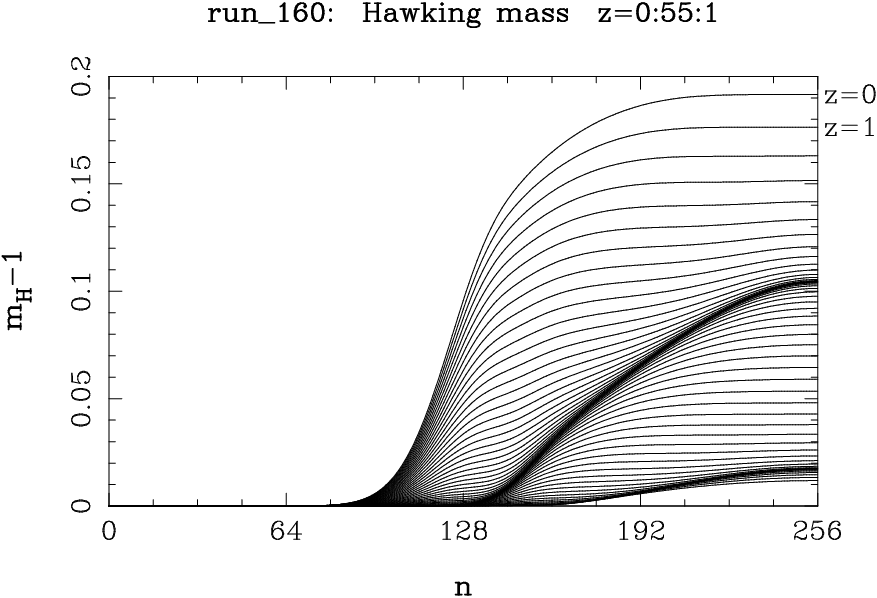}}
     (b) \resizebox{0.4\textwidth}{!}{\includegraphics{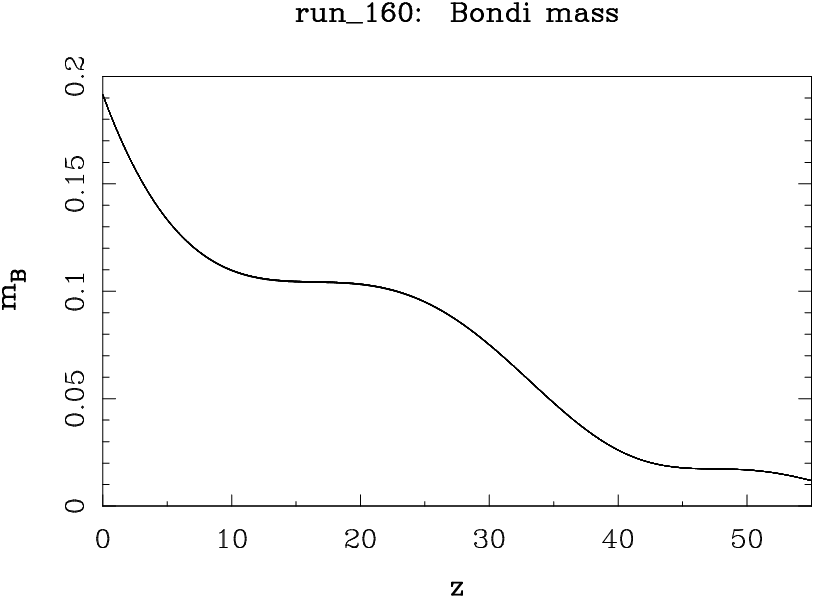}}
  \caption{Mass functions: (a) Hawking mass for times $z=0,1,\ldots,55$;
                           (b) Bondi mass. }
  \label{fig:mH}
\end{figure}

The Hawking mass 
\begin{equation} \label{mH}
  m_{H}(\Sigma) =   \sqrt{\frac{\textrm{area}(\Sigma)}{16\pi}}
  \left(1-\frac{1}{2\pi}\oint_{\Sigma}\rho_{NP}\mu_{NP}\,dv_{\Sigma}\right)
\end{equation}
of a 2-surface $\Sigma$ reduces in the NQS gauge to
\begin{equation}
        m_H(z,r) =  \half r\left( 1 - \frac{1}{8\pi} \oint_{S^2} HJ\right).
        \label{def:mH}
\end{equation}
$m_{H}(z,r)$ provides an easily computed quantity representing the 
``quasi-local'' mass contained within the sphere $(z,r)$, and has 
asymptotic limit equal to the Bondi mass
\begin{equation}
        m_B(z) = \lim_{r\to\infty} m_H(r,z).
        \label{def:mB}
\end{equation}
The Bondi mass is easily computed numerically, by 
$m_{B}(z)=m_{H}(n=n_{\infty},z)$.  Figure \ref{fig:mH}(a) shows the 
Hawking mass plotted against the radial coordinate, for times 
$z=0,1,\ldots,60$.  There are several features of interest in this 
plot: the limit Bondi mass (Figure \ref{fig:mH}(b)) decays in time, 
reflecting the Trautman-Bondi mass loss formula \bref{eq:dzmB}; the 
energy is radiated in bursts, reflecting near-linear behaviour 
dominated by pure $l=2$ modes; the Hawking and Bondi masses decay to 
background black hole mass $\mschw=1$ at late times, suggesting that 
in this example, almost all the gravitational radiation has been 
scattered to $\scri^{+}$ and essentially none will be absorbed by the 
black hole; and finally, the rapidly growing feature about $n=20$ at 
late times in Figure \ref{fig:beta}, does not affect the Hawking mass.  

\begin{figure}[h!tb]
 \centering
    (a) \resizebox{0.4\textwidth}{!}{\includegraphics{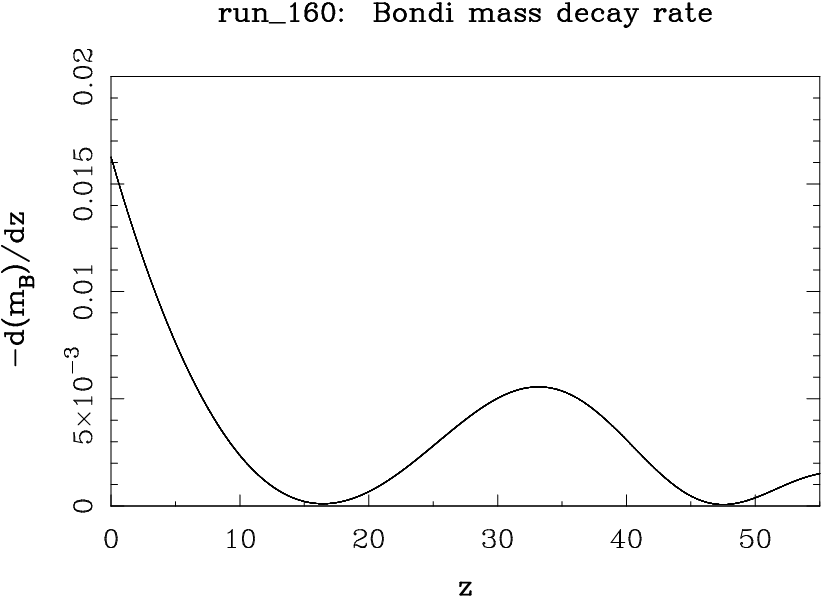}}
    (b) \resizebox{0.4\textwidth}{!}{\includegraphics{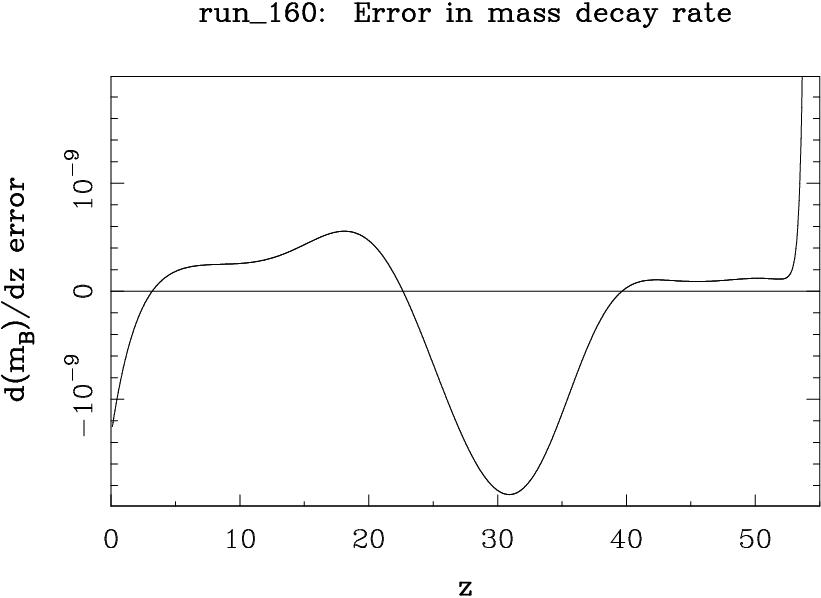}}
   \caption{The Trautman-Bondi mass loss formula as test of 
           numerical accuracy at $r=\infty$: (a)~Bondi mass decay 
           rate; (b)~error in the mass decay formula, given by ${\rm 
           LHS\bref{eq:dzmB}}-{\rm RHS\bref{eq:dzmB}}$.  }
   \label{fig:massloss}
\end{figure}

The Trautman-Bondi mass loss formula 
\cite{Trautman58a,Trautman58b,Bondi60,Bondi62}
\begin{equation}
        \frac{d}{dz}m_{B}(z) = {}- \frac{1}{16\pi} 
        \lim_{r\to\infty}\oint_{S^{2}(z,r)} H|K|^{2}.
\label{eq:dzmB}
\end{equation}
provides another test of the geometric consistency of the solution, 
particularly near null infinity.  By comparing the numerical 
derivative $dm_{B}/dz$ with the computed value of the right hand side 
(evaluated at $n=n_{\infty}$), we may construct the error 
$\frac{d}{dz}m_B-\textrm{RHS(\ref{eq:dzmB})}$.  
Figure \ref{fig:massloss}(b) plots this error against time $z$,  
suggesting that the asymptotic ($r=\infty$) fields of \texttt{run\_160} 
are accurate {}to about $0.00001\%$.

\bibliography{RAB,Energy,PDE,GR,NumericalAnalysis,NumericalGR,Characteristic,Sphere,WinicourEtal}
\bibliographystyle{plain}

\end{document}